\documentclass[onecolumn,english,notitlepage,nofootinbib]{revtex4-1}

\usepackage[T1]{fontenc}
\usepackage{verbatim}
\usepackage{amsmath,bm}
\usepackage{amssymb}
\usepackage{graphicx,color}

\makeatletter

\newcommand{\D}{\mathrm{d}}
\newcommand{\e}{\mathrm{e}}
\newcommand{\half}{\frac{1}{2}}
\newcommand{\vecr}{{\bf r}}

\newcommand{\be}{\begin{equation}}
\newcommand{\ee}{\end{equation}}
\newcommand{\bea}{\begin{eqnarray}}
\newcommand{\eea}{\end{eqnarray}}

\newcommand{\cA}{\mathcal{A}}

\newcommand{\cF}{\mathcal{F}}

\newcommand{\cP}{\mathcal{P}}

\newcommand{\cS}{\mathcal{S}}

\newcommand{\qq}{\begin{equation}}
\newcommand{\qqq}{\end{equation}}
\newcommand{\al}{\begin{aligned}}
\newcommand{\all}{\end{aligned}}

\newcommand{\bfJ}{{\bf J}}

\newcommand{\bfr}{{\bf r}}

\newcommand{\p}{\partial}

\newcommand{\A}{{\text{\tiny A}}}

\newcommand{\E}{{\text{\tiny E}}}
\newcommand{\R}{{\rm R}}

\newcommand{\br}{\textbf{r}}

\renewcommand{\log}{\ln}
\providecommand{\avg}[1]{\left \langle #1 \right \rangle}

\providecommand{\brt}[1]{\left [ #1 \right]}

\providecommand{\f}[2]{\frac{#1}{#2}}

\begin{document}
\title{Stochastic Hydrodynamics of Complex Fluids: Discretisation and Entropy Production}

\author{Michael E. Cates $^{1}$, \'Etienne Fodor $^{2}$, Tomer Markovich $^{3*}$, Cesare {Nardini} $^{4,5}$ and Elsen {Tjhung} $^{6,7}$}

\affiliation{
	$^{1}$ Department of Applied Mathematics and Theoretical Physics,  
	Centre for Mathematical Sciences, University of Cambridge,
	Wilberforce Road, \mbox{Cambridge CB3 0WA, UK} \\
	$^{2}$ Department of Physics and Materials Science, University of Luxembourg, \mbox{L-1511 Luxembourg, Luxembourg}\\
	$^{3}$ Center for Theoretical Biological Physics, Rice University, Houston, TX 77005, USA\\
	$^{4}$ Service de Physique de l'Etat Condens\'e, {CEA}, CNRS Universit\'e Paris-Saclay, CEA-Saclay, \mbox{91191 Gif-sur-Yvette, France}\\
	$^{5}$ {Laboratoire de Physique Th\'eorique de la Mati\`ere Condens\'ee, Sorbonne Universit\'e, CNRS, 75005 Paris, France} 	\\
	$^{6}$ Department of Physics, University of Durham, Science Laboratories, South Road,
	Durham DH1 3LE, UK\\
	$^{7}$  School of Mathematics and Statistics, The Open University, Walton Hall, Milton 	Keynes, MK7 6AA, UK \\ \hfill\\
	$^{*}$  {\rm Corresponding author: tm36@rice.edu} \hfill
}

\begin{abstract}
Many complex fluids can be described by continuum hydrodynamic field equations, to which noise must be added in order to capture thermal fluctuations. In almost all cases, the resulting coarse-grained stochastic partial differential equations carry a short-scale cutoff, which is also reflected in numerical discretisation schemes. We draw together our recent findings concerning the construction of such schemes and the interpretation of their continuum limits, focusing, for simplicity, on models with a purely diffusive scalar field, such as `Model B' which describes phase separation in binary fluid mixtures.  We address the requirement that the steady-state entropy production rate (EPR) must vanish for any stochastic hydrodynamic model in a thermal equilibrium. Only if this is achieved can the given discretisation scheme be relied upon to correctly calculate the nonvanishing EPR for `active field theories' in which new terms are deliberately added to the fluctuating hydrodynamic equations that break detailed balance. To compute the correct probabilities of forward and time-reversed paths (whose ratio determines the EPR), we must make a careful treatment of so-called `spurious drift' and other closely related terms that depend on the discretisation scheme. We show that such subtleties can arise not only in the temporal discretisation (as is well documented for stochastic ODEs with multiplicative noise) but also from spatial discretisation, even when noise is additive, as most active field theories assume. We then review how such noise can become multiplicative via off-diagonal couplings to additional fields that thermodynamically encode the underlying chemical processes responsible for activity. In this case, the spurious drift terms need careful accounting, not just to evaluate correctly the EPR but also to numerically implement the Langevin dynamics itself.
\end{abstract}

\maketitle

\section{Introduction}

Numerous complex fluid systems can be described by continuum equations formulated at the hydrodynamic level. This reflects the fact that their important structure and dynamics arises at a mesoscopic scale not a molecular one. Examples include theories of flowing liquid crystals described by vector or tensor order parameters~\cite{Beris-Edwards-book, deGennes-Prost-book}, and~those of partially miscible binary fluid mixtures, described by a conserved scalar composition variable~\cite{CatesTjhungJFM}. The~latter can undergo phase separation via a combination of diffusive motion and fluid flow, for~which the canonical model is called Model H in the classification of Hohenberg and Halperin~\cite{hohenberg1977theory}. An~important special case of Model H, in~which the fluid velocity is set to zero so that phase separation proceeds by diffusion only, is called Model B. The~latter describes various physical processes in complex fluids, such as Ostwald ripening of emulsion droplets, where the coupling between diffusion and fluid flow is~unimportant.

These hydrodynamic-level descriptions are often first encountered as deterministic equations of motion. This is sometimes sufficient, for~example, in Ostwald ripening of emulsions where large droplets grow at the expense of small ones via {\em deterministic} diffusive fluxes. However, there are many other processes in binary fluids (and also liquid crystals), ranging from droplet nucleation to dynamics near critical points, where the stochasticity of the continuum models must be retained so as to maintain a faithful description of thermal fluctuations. Note that this is even true of single-phase fluids whose true quiescent state involves a Boltzmann distribution for the velocity field ${\bf v}({\bf r})$ 
, not the state of zero velocity predicted by the Navier--Stokes equation in the absence of forcing. As~first shown by Landau and Lifshitz, this is fixed by adding a fluctuating thermal stress to the Navier--Stokes equation \cite{Landau-Lifshitz}. The~resulting thermal fluctuations in the fluid then impart Brownian motion to any colloidal particle suspended in it, without~the need for a separate Langevin force on the~colloid.

In the hydrodynamic modelling of complex fluids, it is therefore important to be able to handle thermal noise terms correctly, both at a conceptual level in the continuum and~when creating discrete implementations of the continuum equations for use in computer simulation studies. The~first of these tasks poses technical challenges of surprising complexity, which can only be resolved by studying the discretisation issue. The~reason for this is simple: adding noise converts the PDEs of deterministic complex fluid models into Stochastic PDEs (SPDEs), which, in general, have no mathematical meaning without some sort of cutoff at short scales. (In a few favourable cases, meaning has been restored directly at the continuum level by a procedure that effectively constructs the renormalization group and the continuum limit simultaneously~\cite{Hairer}.)

In terms of physical modelling, the~existence of a cutoff is unproblematic: continuum descriptions, such as the Beris--Edwards equations for liquid crystals or Models H and B for binary fluids, only hold at scales larger than the molecular one. Mathematically, however, once noise is included, the~cutoff can infiltrate the continuum models in unexpected ways. For~example, we will find below that trying to work directly in the continuum limit gives in the equations under study undefined mathematical objects, such as $\delta({\bf 0})$---the Dirac delta-function evaluated at zero argument. This is symptomatic of a quantity that diverges as the cutoff becomes small. Moreover, we know from equilibrium statistical physics that a particular quantity of interest may or may not depend on the cutoff according to details of the model. For~example, if~a scalar-order parameter field has Gaussian fluctuations at wavenumber ${\bf q}$, $\langle |\phi_{\bf q}|^2\rangle = G^{-1}(q)$, then the corresponding real-space variance $\langle|\phi({\bf r})|^2\rangle$ either remains finite or blows up with the cutoff according to the convergence at high $q$ of $\int G^{-1}(q) \D{\bf q}$. This real-space variance is a legitimate object of enquiry. However, hydrodynamic descriptions such as Model H and B effectively expand $G$ as a low-order polynomial in $q$ on~the basis that the high $q$ behaviour is not important.
For this reason, it is unwise to assume that the continuum limit of the models studied by physicists always make~sense.

Turning  from that conceptual issue to the more practical one of numerically discretising the hydrodynamic equations of a thermal complex fluid, there emerges a crucial requirement for the treatment of noise that creates further surprising traps for the unwary. This is the requirement that the discretised equations respect the principle of detailed balance. Put differently, if~one sets up a numerical model for a complex fluid and calculates its entropy production rate (EPR) in a steady state of thermal equilibrium, the~EPR should vanish. We will see below that there are various different ways in which numerical analyses can fail this~test. 

One setting in which the issue of entropy production comes to the fore is in the study of {\em active field theories
}~\cite{CatesLesHouches}.  These are stochastic hydrodynamic models intended to describe active complex fluids whose microscopic components are driven by an internal power supply. Examples of such active fluids include suspensions of motile bacteria and~of autophoretic colloidal particles with asymmetric surface chemistry that catalyses a chemical reaction, creating chemical gradients that drive the colloids forward. The~study of active matter has exploded into a field whose detailed discussion would take us far beyond the topic of this paper; see~\cite{Marchetti2013}.  For~the present purposes, we can regard active field theories as extensions of the stochastic hydrodynamic equations for complex fluids in which detailed balance is {\em deliberately broken} by the inclusion of new terms that do this, usually at the lowest possible order in the expansion in order parameter fields and their~gradients. 

A strategy we have recently developed in studying such active field theories is to quantify their mesoscopic irreversibility by calculating the steady-state EPR directly at the level of the fluctuating order parameter field dynamics~\cite{Nardini2017,Nardini2018,Borthne2020,Markovich2021}. This quantity is best-called the {\em informatic} EPR or IEPR~\cite{FodorJackCates2021}: it makes no attempt to capture all the microscopic irreversibility or heat flows associated with the particle motions underlying the coarse-grained, hydrodynamic SPDEs. Instead, the IEPR is computed informatically from forward and reverse path probabilities using the tools of stochastic thermodynamics~\cite{Seifert2012} applied to the SPDEs themselves.
These tools have also found applications in active matter systems such as biochemical signalling~\cite{Horowitz-arxiv-2021}, mechanosensory processes~\cite{Parrondo-njp-2021} and~bacterial motion~\cite{Bo-prx-2019}.
We have further extended these ideas and embedded a large class of active field theories in a thermodynamically consistent setting that accounts for their driving mechanism, in~which case, the~irreversibility of the enlarged system capture the actual rate of heat production.
In our studies of active field theories, we have found interesting physics to be laid bare when one considers the way the IEPR (and the heat rate) depends on the spatial configuration of the system and also the way different contributions to it (e.g., bulk or interfacial) scale with the noise level. To~address these issues by computer simulation, it is clearly crucial to have a numerical implementation in which the calculated entropy production arises solely by virtue of the active, detailed-balance-breaking terms, unpolluted by any failure of the numerical discretisation scheme to respect detailed balance even in thermal~equilibrium.

Accordingly, in~our recent studies of active field theories, we have been forced to carefully consider the conceptual and discretisation issues for the stochastic hydrodynamics of complex fluids generally. We have found that, beyond~a few important contributions such as~\cite{Lau2007, Lecomte2017}, these issues are not widely discussed in the literature accessible to \mbox{physicists---especially }not in relation to entropy production and its numerical evaluation. Thus far, our own results on these topics have been presented only incidentally, if~at all, in~technical appendices and side remarks in papers on how active hydrodynamic models actually behave~\cite{Nardini2017,Nardini2018,Borthne2020,Markovich2021}. We attempt here a coherent perspective on these issues. For~simplicity, our main focus is on Model B and its active counterparts, in~which the sole order parameter is a scalar field and the only dynamics is diffusive. Indeed, between~here and the concluding section, we say nothing of the wider class of complex fluid models containing vector and tensor order parameters (for liquid crystals) or even a coupling to fluid flow (for a scalar field, Model H). We emphasize, however, that the conceptual and discretisation issues addressed here apply, in~varying degrees, to~all these other~cases.

The rest of this paper is structured as follows. To~set the stage, Section~\ref{sec:1D} reviews the questions of discretisation and spurious drift for a single particle Langevin equation with multiplicative noise, discussing also the Fokker--Planck equation, path integrals, and~entropy production in this simplified setting before addressing the stochastic calculus for finitely many degrees of freedom. This establishes a core set of ideas that are utilized subsequently for the continuum case. In~Section~\ref{sec:additive}, we turn to continuous fields, focusing on the case of (active) Model B where the noise is additive rather than multiplicative, and~show how the spatial discretisation must be carefully handled to avoid erroneous evaluation of the (informatic) entropy production. We focus on finite difference schemes, as~opposed to spectral ones, for~spatial discretisation because, besides~being widely used, this approach offers the most direct way to illuminate problems with the continuum limit. In~Section \ref {sec:Etienne-Tomer}, we consider how to embed an active field theory within an enlarged description that is thermodynamically consistent in the sense that it accounts for heat flow (caused, in~this instance, by~chemical reactions that drive the system microscopically) at the level of linear irreversible thermodynamics. We review how this generically leads to multiplicative noise even where none was previously present and~describe the further conceptual and discretisation problems arising from this. Finally, in~Section~\ref{sec:conclusion}, we offer some brief concluding~remarks.

\section{Stochastic Thermodynamics of~Particles \label{sec:1D}} 

In this section, we establish some basic concepts concerning stochastic differential equations and thermal motion, starting in the context of a single particle and then turning to the case with several degrees of~freedom.

\subsection{Langevin~Equation}

Let us consider a colloidal particle suspended inside a viscous solvent in one dimension. 
The solvent acts as a heat bath for the particle with temperature $T$, and the particle is assumed to be in thermal equilibrium with the heat bath at all times.
Let us denote $x(t)$ the stochastic trajectory of the centre of mass of the particle.
The equation of motion for the particle is then given by the overdamped Langevin equation:
\begin{equation}
\frac{dx}{dt} = \underbrace{-\Gamma(x)U'(x)+\nu_{a}(x)}_{f(x)} + \underbrace{\sqrt{2D(x)}}_{g(x)} \eta(t) \, , \label{eq:1D-Langevin}
\end{equation}
where $U(x)$ is an external potential (provided, {e.g.}, by~an optical trap), and~$\eta(t)$ is a Gaussian white noise with zero mean $\left\langle \eta(t)\right\rangle =0$
and (Dirac) delta-function correlation $\left\langle \eta(t)\eta(t')\right\rangle =\delta(t-t')$. 
Note that $x(t)$ is a stationary process since the potential $U$ does not explicitly depend on time.
In Equation~(\ref{eq:1D-Langevin}), we neglect the inertia of the particle, which is valid if the Reynolds number is much smaller than unity; 
$\Gamma(x)$  is the mobility or the inverse of the friction coefficient. 
(For a spherical particle, $\Gamma=1/(6\pi\eta R)$, where $\eta$ is the viscosity of the solvent and $R$ is the radius of the particle.) 
In this example, we also allow the mobility $\Gamma(x)$ to vary locally in space. 
$D(x)$ in Equation~(\ref{eq:1D-Langevin}) is the diffusion coefficient or the noise strength. 
The noise strength $D(x)$, the~mobility $\Gamma(x)$, and~the solvent temperature $T$ are all related via the Stokes--Einstein relation, which is a direct consequence of the fluctuation-dissipation theorem (FDT):
$D(x)=\Gamma(x)T$ (note that we work in units of $k_{B}=1$). 
Since the mobility, and~hence the diffusion constant, vary locally in space,  the~noise in Equation (\ref{eq:1D-Langevin}) is  multiplicative.
Since the noise is multiplicative, the~Langevin equation as written in Equation (\ref{eq:1D-Langevin}) is ambiguous, 
unless we specify how we discretise the dynamics in time.
Depending on how we do so, we may need to include the {\em spurious drift} term $\nu_{a}(x)$ in Equation (\ref{eq:1D-Langevin}) 
to recover Boltzmann distribution in the steady state. The~`spurious drift' terminology is conventional but~may be confusing: the term $\nu_{a}(x)$ arises in effect because the noise, depending on the discretisation scheme used, might or might not still have zero average. (For instance, if~the noise variance increases with $x$ and is evaluated mid-step, then random steps in the positive $x$ direction are larger than those towards negative $x$.)
Finally, to simplify the notation, we shall also define:
\begin{align}
f(x) &= -\Gamma(x)U'(x) + \nu_{a}(x) \label{eq:1D-f} \, ,\\
g(x) &= \sqrt{2D(x)} \, . \label{eq:1D-g}
\end{align}

\subsection{Discretised Langevin~Equation}

Let us discretise the time into $t_{n}=t_{0}+n\Delta t$, where $n=0,1,2,\dots,N$.
The Dirac delta function correlation in the continuous noise $\left\langle \eta(t)\eta(t')\right\rangle =\delta(t-t')$ can be regularized into a Kronecker delta 
$\left\langle \eta_{m}\eta_{n} \right\rangle = \delta_{mn}/\Delta t$.
The discretised Langevin equation is then given by:
\begin{equation}
\Delta x_{n} = x_{n+1} - x_{n} = f(x_{n+a})\Delta t + g(x_{n+a})\xi_{n}\sqrt{\Delta t} \, , \label{eq:1D-Langevin-discrete}
\end{equation}
where $\{\xi_{0},\xi_{1},\dots\xi_{N-1}\}$ are a set of independent Gaussian random variables with zero mean, $\left\langle \xi_{n} \right\rangle = 0$,
and Kronecker delta correlation $\left\langle \xi_{m}\xi_{n} \right\rangle = \delta_{mn}$.
In (\ref{eq:1D-Langevin-discrete}), $a\in[0,1]$ is the discretisation parameter, 
which tells us when during the timestep we should evaluate the particle position $x$ for the purpose of sampling the noise (whose variance is, we recall,  $x$-dependent). Thus, $a=0$ corresponds to the It\^{o} choice (initial postion), 
$a=1/2$ corresponds to Stratonovich (midpoint position), and~
$a=1$ corresponds to anti-It\^{o} (final position). 
Now, using the mean value theorem $x_{n+a} = x_{n} + a\Delta x_{n}$, we write Equation~(\ref{eq:1D-Langevin-discrete}) as:
\begin{equation}
\Delta x_{n} = f(x_{n})\Delta t + g(x_{n})\xi_{n}\sqrt{\Delta t} + a g(x_{n})g'(x_{n})\xi_{n}\xi_{n}\Delta t + \mathcal{O}(\Delta t^{3/2}) \, . \label{eq:1D-Langevin-discrete-2}
\end{equation}

In order to derive the Fokker--Planck equation below, we first need to compute the first and second moment of $\Delta x_{n}$:
\begin{align}
\left\langle \Delta x_{n} \right\rangle  &= f(x_{n})\Delta t + a g(x_{n})g'(x_{n})\Delta t + \mathcal{O}(\Delta t^{3/2}) \, ,\label{eq:1D-Dx} \\
\left\langle \Delta x_{n}\Delta x_{n} \right\rangle  &= g(x_{n})^{2}\Delta t + \mathcal{O}(\Delta t^{3/2}) \, . \label{eq:1D-Dx-Dx}
\end{align}

\subsection{Fokker--Planck~Equation}

Let us denote $P(x,t|x_{0},t_{0})dx$ to be the probability of finding the particle at $[x,x+dx]$ at time $t$, given that it was at $x_{0}$ at time $t_{0}$, 
where $t_{0}<t$. 
The time evolution of this probability density function is given by Kramers--Moyal expansion (see~\cite{VanKampenBook} for derivation):
\begin{eqnarray}
\nonumber P(x,t+\Delta t)-P(x,t) &=& - \frac{\partial}{\partial x} \left[ P(x,t) \left\langle \Delta x(t) \right\rangle \right] \\
&+& \frac{1}{2}\frac{\partial^{2}}{\partial x^{2}} \left[ P(x,t) \left\langle \Delta x(t)\Delta x(t) \right\rangle \right]
+ \mathcal{O}(\Delta t^{3/2}) \, ,
\end{eqnarray}
where $\Delta x(t)=x(t+\Delta t)-x(t)$. 
Substituting Equations (\ref{eq:1D-Dx}--\ref{eq:1D-Dx-Dx}) into the equations above and taking the limit $\Delta t\rightarrow0$, we obtain:
\begin{equation}
\frac{\partial P(x,t)}{\partial t} = -\frac{\partial}{\partial x} \left[ \left\{ f(x) + a g(x)g'(x) \right\} P(x,t) \right]
+ \frac{1}{2}\frac{\partial^{2}}{\partial x^{2}} \left[ g(x)^{2} P(x,t) \right] \, . \label{eq:1D-Fokker--Planck}
\end{equation}

We can also write this as a continuity equation $\partial P/\partial t = -\partial J/\partial x$, where the probability current is given by:
\begin{eqnarray}
\nonumber J(x,t) &=& -\Gamma(x)U'(x)P(x,t) + \nu_{a}(x)P(x,t) + a D'(x)P(x,t) \\
&-& D'(x)P(x,t) - D(x)P'(x,t) \, . \label{eq:1D-J}
\end{eqnarray}
For an equilibrium system, which is the case in our example, the~probability current should be equal to~\cite{Lau2007,Basu2008}:
\begin{equation}
J(x,t) = -\Gamma(x)U'(x)P(x,t) - D(x)P'(x,t) \, . \label{eq:1D-J-eq}
\end{equation}
Together with FDT, $D(x)=\Gamma(x)T$, the~probability current from Equation (\ref{eq:1D-J-eq}) will guarantee Boltzmann distribution in the steady state: 
$P(x,t\rightarrow\infty) \propto \e^{-U(x)/T}$.
Comparing Equation (\ref{eq:1D-J-eq}) to Equation (\ref{eq:1D-J}), we thus require the spurious drift to be
\begin{equation}
\nu_{a}(x) = (1-a)D'(x) \, . \label{eq:1D-spurious}
\end{equation}

In the case of additive noise, where $D$ and $\Gamma$ are constant, the~spurious drift is always zero. 
In the case of multiplicative noise, where $D(x)$ and $\Gamma(x)$ vary in space, the~spurious drift can be made to vanish only by choosing the anti-It\^{o} discretisation $a=1$. Generally speaking, numerical strategies are simplest for It\^{o} ($a=0$), whose update statistics depend only on the state at the start of the timestep in which the update is to occur. On~the other hand, the~Stratonovich discretisation ($a=1/2$) has some desirable properties in relation to temporal reversibility, see Section~\ref{sec:Elsen-EP}.
Moreover, as~we will see in Equation (\ref{eq:d-fg}) below, setting $a=1$ does not eliminate all spurious drift terms in higher dimensions, where such terms remain a generally unavoidable~feature.

\subsection{Path Integral~Formalism}

The Fokker--Planck equation in Equation (\ref{eq:1D-Fokker--Planck}) is usually rather difficult to solve when generalising to higher dimensions. 
In many situations ({e.g.}, when calculating the entropy production rate), it is often easier to work with the path~probability.

\subsubsection{Transition~Probability}

Suppose that our particle is initially at $x_{n}$ at time $t_{n}$.
For a given noise realization $\xi_{n}$, 
the position of the particle $x_{n+1}$ in the next timestep $t_{n+1}$ is given by the discretised Langevin Equation~(\ref{eq:1D-Langevin-discrete}), in~which 
$\xi_{n}$ is a Gaussian random variable with probability \mbox{density function:}
\begin{equation}
P_\xi(\xi_{n}) \,d\xi_{n}= \frac{1}{\sqrt{2\pi}} \e^{-\frac{1}{2}\xi_{n}^{2}}\, d\xi_{n} \, . \label{eq:1D-Gaussian}
\end{equation}
We can then substitute Equation~(\ref{eq:1D-Langevin-discrete}) into Equation~(\ref{eq:1D-Gaussian}) 
to obtain the probability of finding the particle at $[x_{n+1},x_{n+1}+dx_{n+1}]$ at time $t_{n+1}$, given that it was at $x_{n}$ at the previous timestep $t_{n}$:
\begin{align}
P(x_{n+1}|x_{n})dx_{n+1} &= \frac{1}{\sqrt{2\pi}}
\e^{-\frac{\Delta t}{2g(x_{n+a})^{2}} \left[ \frac{x_{n+1} - x_{n}}{\Delta t} - f(x_{n+a}) \right]^{2}}
\left| \frac{d\xi_{n}}{dx_{n+1}} \right|
dx_{n+1} \, . \label{eq:1D-transition-prob}
\end{align}
Note that the Jacobian $\left| d\xi_{n}/dx_{n+1} \right|$ is inserted when we change the random variable from $\xi_{n}$ to $x_{n+1}$. 
To find the Jacobian, we first express $\xi_{n}$ as a function of $x_{n+1}$ from Equation~(\ref{eq:1D-Langevin-discrete}) to obtain:
\begin{equation}
\xi_{n} = \frac{1}{g(x_{n}+a(x_{n+1}-x_{n})) \sqrt{\Delta t}} \left[ x_{n+1} - x_{n} - f(x_{n} + a(x_{n+1} - x_{n})) \Delta t \right] \, ,
\end{equation}
where we have used the mean value theorem $x_{n+a} = x_{n} + a(x_{n+1}-x_{n})$ again. 
Taking a derivative with respect to $x_{n+1}$, we then obtain:
\begin{equation}
\frac{d\xi_{n}}{dx_{n+1}} = \frac{1}{g(x_{n+a})\sqrt{\Delta t}}
\left\{ 1 - \Delta ta f'(x_{n+a}) 
- \Delta ta\frac{g'(x_{n+a})}{g(x_{n+a})}
\left[ \frac{x_{n+1} - x_{n}}{\Delta t} - f(x_{n+a}) \right] 
\right\} \, . \label{eq:1D-Jacobian}
\end{equation}

We want to exponentiate the terms inside the curly bracket; however, we note that $x_{n+1}-x_{n}\sim\sqrt{\Delta t}$, so we cannot conduct this directly. 
Instead, we shall use the following Taylor expansion~\cite{Lecomte2017}, for~some constant $C\sim\Delta t^{0}$:
\begin{align}
\e^{C\Delta x_{n}} &= 1 + C\Delta x_{n} + \frac{1}{2}C^{2} \underbrace{\Delta x_{n}\Delta x_{n}}_{g^{2}(x_{n+a})\Delta t} + \mathcal{O}(\Delta t^{3/2}) \, ,\\
\Rightarrow 
\e^{C\Delta x_{n} - \frac{1}{2}C^{2}g^{2}(x_{n+a})\Delta t} &= 1 + C\Delta x_{n} + \mathcal{O}(\Delta t^{3/2}) \, . \label{eq:1D-exponential}
\end{align}
Here, we have approximated $\Delta x_{n}\Delta x_{n}\simeq g^{2}(x_{n+a})\Delta t$, which is valid for small $\Delta t$, 
\emph{c.f.} Equation~(\ref{eq:1D-Dx-Dx}) and~\cite{Lecomte2017}. 
The Jacobian can then be exponentiated as follows:
\begin{equation}
\frac{d\xi_{n}}{dx_{n+1}} = \frac{1}{g\sqrt{\Delta t}}
\exp\left\{ -\Delta ta f' - \Delta ta\frac{g'}{g}\left(\frac{x_{n+1}-x_{n}}{\Delta t} - f \right)
-\frac{\Delta t}{2}a^{2}g'^{2}
+\mathcal{O}(\Delta t^{3/2})
\right\} \, , \label{eq:1D-Jacobian-1}
\end{equation}
where $f$, $g$, $f'$, and~$g'$ are evaluated at $x_{n+a}$.
Substituting Equation (\ref{eq:1D-Jacobian-1}) into Equation (\ref{eq:1D-transition-prob}), we obtain:
\begin{eqnarray}
\nonumber P(x_{n+1}|x_{n})  = \frac{1}{\sqrt{2\pi g^{2}\Delta t}}
\exp\Bigg\{ &-&\frac{\Delta t}{2g^{2}} \left[ \left( \frac{x_{n+1}-x_{n}}{\Delta t} - f \right)^{2} + 2a gg' \left( \frac{x_{n+1}-x_{n}}{\Delta t} - f \right)\right] \\
&-& \Delta t\left( \half a^{2}g'^{2} + a f' \right) +\mathcal{O}(\Delta t^{3/2})\Bigg\} \, ,
\end{eqnarray}
where $f$, $g$, $f'$, and~$g'$ are again evaluated at $x_{n+a}$.
Finally after completing the square, we obtain the transition probability:
%
\begin{eqnarray}
\nonumber& &P(x_{n+1}|x_{n}) = \frac{1}{\sqrt{4\pi D(x_{n+a})\Delta t}} \\
\nonumber& &\times \exp\Bigg[-\Delta t \left\{ \frac{1}{4D(x_{n+a})} \left[\frac{x_{n+1}-x_{n}}{\Delta t} - f(x_{n+a}) + a D'(x_{n+a})\right]^{2}
+ a f'(x_{n+a}) \right\} \\
& &+ \mathcal{O}(\Delta t^{3/2}) \Bigg] \, . \label{eq:1D-transition-prob-1}
\end{eqnarray}

Below, we will often use a shorthand notation whereby the $\mathcal{O}(\Delta t^{3/2})$ term is implicit in expressions such as~this.

\subsubsection{Path~Integral}

Suppose that, initially, the particle is at $x_{0}$ at time $t_{0}$.
What is the probability that we find the particle at $[x_{N},x_{N}+dx_{N}]$ at time $t_{N}$? 
This probability can be written as (Chapman--Kolmogorov equation):
\begin{equation}
P(x_{N}|x_{0}) = \int\D x_{1} \int\D x_{2} \dots \int\D x_{N-1} \, P(x_{N}|x_{N-1})P(x_{N-1}|x_{N-2}) \dots P(x_{1}|x_{0}) \, . 
\end{equation}
Substituting the transition probability from Equation (\ref{eq:1D-transition-prob-1}) into the equation above, we obtain:
\begin{equation}
P(x_N|x_0) = \int \D x_1 \int \D x_2 \dots \int\D x_{N-1} \, \mathcal{N}\{x_n\} \, \e^{-\mathcal{A}\{x_n\}}  \, ,\label{eq:1D-path-integral}
\end{equation}
where $\mathcal{A}\{x_{n}\}$ is called the dynamical action (Onsager--Machlup action)~\cite{Lecomte2017,Lau2007}:
\begin{equation}
\mathcal{A}\{x_{n}\} = \sum_{n=0}^{N-1}\Delta t
\left\{ \frac{1}{4D(x_{n+a})} \left[ \frac{x_{n+1}-x_{n}}{\Delta t} - f(x_{n+a}) + a D'(x_{n+a})\right]^{2}
+ a f'(x_{n+a})
\right\} \, , \label{eq:1D-action}
\end{equation}
and $\mathcal{N}\{x_n\}$ is some normalization prefactor, which is constant for additive noise:
\begin{equation}
\mathcal{N}\{x_n\} = \prod_{n=0}^{N-1} \frac{1}{\sqrt{4\pi D(x_{n+a})\Delta t}} \, . \label{eq:1D-N}
\end{equation}

In the limit $\Delta t\rightarrow0$, Equation~(\ref{eq:1D-path-integral}) becomes a path integral, 
	{i.e.}, we sum over all possible trajectories $\{x_{n}|n=0,1,2,\dots,N\}$,
	each with a weight or path probability 
	$\mathcal{P}\{x_{n}\} = \mathcal{N}\{x_n\}\,\e^{-\mathcal{A}\{x_{n}\}}$.
	Here, $f(x)=-\Gamma(x)U'(x)+(1-a)D'(x)$ as in \mbox{Equations~(\ref{eq:1D-Langevin}) and (\ref{eq:1D-spurious}).}
	Note that the expressions for $\mathcal{N}$ and $\mathcal{A}$ in Equations (\ref{eq:1D-action}--\ref{eq:1D-N}) are generic for any processes.
	When we invoke FDT below, $D(x)=\Gamma(x)T$, we then assume $\{x_n\}$ to be a stationary process.
	Furthermore, note that for additive noise, where $D$ and $\Gamma$ are constant, the~dynamical action still depends on the discretisation parameter $a$,
	even though the discretised Langevin dynamics does not depend on $a$ anymore. This is important when calculating entropy production via the path probabilities, as~we consider next.

\subsection{Entropy~Production}\label{sec:Elsen-EP}

\begin{figure}[h]
	\centering
	\includegraphics[width=\linewidth]{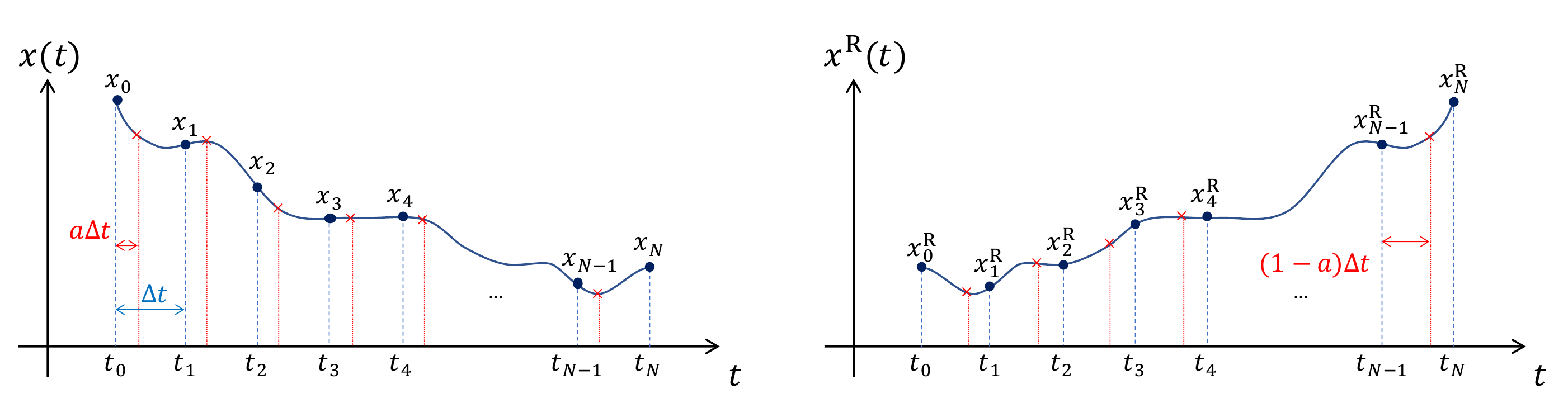}
	\caption{\label{fig:trajectories}
		The backward trajectory $x^\R(t)$ ({\bf right}) is obtained by reflecting the forward trajectory $x(t)$ ({\bf left}) around the vertical line $t=(t_N+t_0)/2$.
		The discretisation parameter $a$ for the forward trajectory ({\bf left}) becomes $1-a$ for the backward trajectory ({\bf right}).
	}
\end{figure}

Consider now a single stochastic trajectory of our overdamped particle, $\{x_{n}|n=0,1,2\dots N\}$, 
which was generated by the discretised Langevin Equation~(\ref{eq:1D-Langevin-discrete}).
A foundational result of stochastic thermodynamics~\cite{Seifert2012,Peliti-book} is that the total heat dissipated from the particle to the environment, as~it moves along this single trajectory, obeys:
\begin{equation}
{\cal Q} = T\ln\frac{\mathcal{P}\{x_{n}\}}{\mathcal{P}^\R\{x_{n}\}} = T\,\Delta S_{m} \, . \label{eq:1D-Q}
\end{equation}
The second equality states that the heat dissipation determines the increase in the entropy of the medium or heat bath supplying the noise, $\Delta S_{m}$. 
In (\ref{eq:1D-Q}), $\mathcal{P}\{x_{n}\}$ is the probability of obtaining this particular trajectory $\{x_{n}|n=0,1,2,\dots N\}$
and 
$\cP^\R\{x_{n}\}$ is the probability of observing the time-reversed trajectory 
$\{x_{n}^{\R}=x_{N-n}|n=0,1,2,\dots N\}$ (see Figure~\ref{fig:trajectories}) under~the same Langevin dynamics (\ref{eq:1D-Langevin-discrete}).
For example, the~chosen trajectory might be a particle going from high to low energy, in~which case 
the time-reversed trajectory is much less probable to be seen under the same forward Langevin dynamics~(\ref{eq:1D-Langevin-discrete}).
Note that to keep the same notation as in the previous literature~\cite{FodorJackCates2021,Fodor2016,Markovich2021}, $\mathcal{P}^\R\{x_{n}\}:=\cP\{x^\R_n\}$ in Equation (\ref{eq:1D-Q}).
Furthermore, note that if the specific trajectory $\{x_n\}$ results from a specific protocol, such as changing one of the parameters inside the potential energy $U(x)$ (which we do not consider in this paper), in~the time-reversed trajectory, we also have to reverse the direction of this protocol~\cite{Jarzynski-review-2011,Parrondo-prl-2011,Seifert2012,Parrondo-njp-2021}.

\unskip

\subsubsection{Evaluation via Discretised~Action}

For the forward trajectory $\{x_n\}$, the~path probability obeys $\mathcal{P}\{x_{n}\}=\mathcal{N}\{x_n\}\, \e^{-\mathcal{A}\{x_{n}\}}$,
where $\mathcal{N}\{x_n\}$ and $\mathcal{A}\{x_{n}\}$ are given in Equations~(\ref{eq:1D-action}) and (\ref{eq:1D-N}), respectively.
For the time-reversed trajectory $\{x^\R_n\}$, the~path probability is given by $\mathcal{P}\{x^\R_{n}\}=\mathcal{N}\{x^\R_n\}\, \e^{-\mathcal{A}\{x^\R_{n}\}}$, where $\mathcal{N}$ and $\mathcal{A}$ are still the same expressions, given in Equations~(\ref{eq:1D-action}) and (\ref{eq:1D-N}), except~that we replace the arguments by $\{x^\R_n\}$.
Let us first calculate the normalization prefactor $\mathcal{N}\{x^\R_n\}$ for the \mbox{reversed trajectory $\{x^\R_n\}$:}
\begin{equation}
\mathcal{N}\{x^\R_n\} = \prod_{n=0}^{N-1} \frac{1}{\sqrt{4\pi D(x^\R_{n+a})\Delta t}} = \prod_{n=0}^{N-1} \frac{1}{\sqrt{4\pi D(x_{n+(1-a)})\Delta t}} \, ,
\label{eq:1D-N-R}
\end{equation}
In the second equality above, we have substituted $x^\R_n=x_{N-n}$.
Comparing Equation (\ref{eq:1D-N}) to Equation (\ref{eq:1D-N-R}), 
the normalization factor $\mathcal{N}$ differs between forward and backward discretised paths and does not cancel in Equation~(\ref{eq:1D-Q}),
unless we choose the Stratonovich discretisation, $a=1/2$.
This is a compelling reason to choose Stratonovich when calculating the heat dissipation or entropy production via Equation (\ref{eq:1D-Q}), and~we do so hereafter. 
In Stratonovich, the~action for the reversed trajectory $\{x^\R_n\}$ is then given by:
\begin{align}
\nonumber\mathcal{A}\{x^\R_{n}\} 
&= \sum_{n=0}^{N-1}\Delta t
\Bigg\{ \frac{1}{4D(x_{n+\frac{1}{2}})} \left[ -\frac{x_{n+1}-x_{n}}{\Delta t} - f(x_{n+\frac{1}{2}}) + \frac{1}{2}D'(x_{n+\frac{1}{2}}) \right]^{2} \\
& + \frac{1}{2} f'(x_{n+\frac{1}{2}}) \Bigg\} \, .
\end{align} 

The heat dissipated is simply the difference between the backward and the \mbox{forward action:}
\begin{align}
{\cal Q} &= T\left( \mathcal{A}^\R\{x_{n}\} - \mathcal{A}\{x_{n}\} \right) \\
\nonumber &= T\sum_{n=0}^{N-1}\frac{\Delta t}{4D(x_{n+\frac{1}{2}})}
\Bigg\{ \left[-\frac{x_{n+1}-x_{n}}{\Delta t} + \Gamma(x_{n+\frac{1}{2}})U'(x_{n+\frac{1}{2}})\right]^{2} \\
\nonumber &\qquad\qquad\qquad\qquad\qquad- \left[\frac{x_{n+1}-x_{n}}{\Delta t} + \Gamma(x_{n+\frac{1}{2}})U'(x_{n+\frac{1}{2}})\right]^{2} \Bigg\} \\
&= -T\sum_{n=0}^{N-1}\frac{\Delta t}{D(x_{n+\frac{1}{2}})} \left(\frac{x_{n+1}-x_{n}}{\Delta t}\right) \Gamma(x_{n+\frac{1}{2}})U'(x_{n+\frac{1}{2}}) \, .
\end{align}
Finally, we apply FDT $D(x)=\Gamma(x)T$ to obtain
\begin{align}
{\cal Q} & = -\underbrace{\sum_{n=0}^{N-1}\Delta t\left(\frac{x_{n+1}-x_{n}}{\Delta t}\right)U'(x_{n+\frac{1}{2}})}_{\text{Stratonovich integral}}
= -\int_{t_{0}}^{t_{N}}\textrm{d}t\,\frac{\textrm{d}x}{\textrm{d}t}  U'(x)
= -\Delta U \, ,\label{eq:1D-first-law}
\end{align}
where $\Delta U  = U(t_N) - U(t_0)$, and since this is a Stratonovich integral, we have used the standard chain rule in the last equality.
Thus, we recover the first law of thermodynamics. 
The Stratonovich integral over $U'(x)\,\textrm{d}x$ in Equation (\ref{eq:1D-first-law}) can, if~desired, be converted to an It\^{o} integral by setting $U'(x_{n+\frac{1}{2}})=U'(x_{n})+\frac{1}{2}U''(x_{n})\Delta x_{n}$ to obtain:
\begin{equation}
{\cal Q} = - \underbrace{\sum_{n=0}^{N-1}\Delta t\left(\frac{x_{n+1}-x_{n}}{\Delta t}\right)U'(x_{n})}_{\text{It\^{o} integral}} 
- \underbrace{\sum_{n=0}^{N-1}\Delta t\,U''(x_{n})D(x_{n})}_{\text{Correction term}}.
\end{equation}
Finally, substituting Equation (\ref{eq:1D-first-law}) back to Equation (\ref{eq:1D-Q}), we may also show that detailed balance is obeyed:
\begin{equation}
\frac{\mathcal{P}\{x_{n}\}}{\cP^\R\{x_{n}\}} = \e^{-\Delta U/T},
\end{equation}
as is indeed required for any system in thermal~equilibrium.

\subsubsection{Non-Equilibrium Steady~State}\label{sec:ness}

We may generalize the above result to non-equilibrium steady states. 
For example, we may imagine applying a non-conservative force $F(x)$ on the particle so that the Langevin equation now reads:
\begin{equation}
\frac{dx}{dt}=\Gamma(x)\left[-U'(x)+F(x)\right]+\nu_{a}(x)+\sqrt{2D(x)}\eta(t) \, ,
\label{eq:1D-driven}\end{equation}
where $D(x)=\Gamma(x)T$. 
In the case of a periodic potential $U(x)$, a~constant external force $F$ 
may give rise to a steady-state current, which indicates a non-equilibrium steady state and thus breaks detailed balance. 
The EPR in the steady-state ensemble is found as follows:
\begin{equation}
\dot{\mathcal{S}} = \lim_{t\rightarrow\infty} \left\langle \frac{1}{t} \ln\frac{\mathcal{P}[x(t')]}{\cP^\R[x(t')]} \right\rangle 
= \frac{\dot{\mathcal{Q}}}{T} \, , \label{eq:1D-entropy}
\end{equation}
where $\mathcal{P}[x(t')]$ is the path probability for some forward trajectory $\{x(t')|t'\in[0,t]\}$ and 
$\cP^\R[x(t')]$ is the path probability for the same trajectory going backwards in time $\{x(t-t')|t'\in[0,t]\}$. 
The angle bracket indicates ensemble averaging or~the average over different noise realizations $\{\eta(t')|t'\in[0,t]\}$.
Note that, since the entropy content of the system is unchanging in the steady state, all the entropy produced within it ends up in the medium or heat bath so that the EPR, which is the rate of change of entropy in the bath $\dot{\mathcal{S}}$, equates to the dissipation rate $\dot{\mathcal{Q}}$  within a factor $T$. 
The notation in (\ref{eq:1D-entropy}) is chosen to connect with subsequent sections and with the previous literature; note, however, that in~\cite{Nardini2017,Nardini2018,Markovich2021}, the un-accented symbol $\mathcal{S}$ is used to denotes the entropy production {\em rate}, which is called $\dot{\mathcal{S}}$ in this paper.

Following the same derivation as above, we can show that the steady-state heat production rate is:
\begin{equation}
\dot{\mathcal{Q}} = T\dot{\mathcal{S}}
= - \left\langle \frac{dU}{dt}\right\rangle + \left\langle F\frac{dx}{dt}\right\rangle 
= \left\langle F\frac{dx}{dt} \right\rangle \, .
\end{equation}
Note that $dU/dt$ is zero in the steady state, on~average. 
Thus, the rate of heat dissipation is equal to the average rate of work conducted by the external force $F$, again consistently with the first~law.

The above results are given in thermodynamic language which ultimately rests on the first law (conservation of energy). However, Equation~(\ref{eq:1D-driven}) is not generically thermodynamically consistent~\cite{Seifert2018,Pietzonka2019}. For~example, one could interpret Equation (\ref{eq:1D-driven}) as describing an active particle, such as a swimming microorganism, for~which the term $\Gamma(x)F(x) = V(x)$ is a spatially varying propulsive velocity. The~$x$ dependence  of $V$ might then have no connection with energetics (that is, $F(x)$ is no longer a mechanical force), reflecting instead a tendency to swim in the positive or negative direction depending on external stimuli, such as an imposed gradient in nutrient levels (for instance, $\Gamma F \propto \partial_xH(x)$ with $H$ a food concentration)~\cite{Murray}.
In such cases, there is no first law behind  Equation (\ref{eq:1D-driven}), and we cannot associate $\ln({\mathcal{P}}/{\cP^\R})$ in Equation~(\ref{eq:1D-entropy}) with energy dissipation or heat production. 
We can nonetheless define an {\em informatic} entropy production rate, or~IEPR, via the first equality only in Equation (\ref{eq:1D-entropy}).
It is this IEPR that we will generalize in Section~\ref{sec:additive} as a tool for quantifying the irreversibility of active field theories. Thereafter, in~Section~\ref{sec:Etienne-Tomer}, we will restore a link with thermodynamics and the first law, under~specific assumptions concerning the near-equilibrium character of the microscopic dynamics responsible for~activity.

\subsection{Stochastic Calculus for $d>1$ Degrees of~Freedom}\label{sec:Elsen-d>1}

Let us consider the general Langevin equation for $d>1$ degrees of freedom in a system with detailed balance. 
We denote the coordinates to be $x_{i}$, where $i=1,2,\dots d$. (This could describe either one particle in $d$ dimensions, or~$N>1$ particles in $d/N$ dimensions.) 
The Langevin equation for $\{x_{i}(t)\}$ is given by
\begin{equation}
\frac{dx_i}{dt} = f_i(\{x_i\}) + g_{ij}(\{x_i\})\eta_j(t) \, ,
\end{equation}
where $\{\eta_{i}(t)\}$ are Gaussian white noises with zero mean $\left\langle \eta_{i}(t) \right\rangle = 0$
and delta-correlations $\left\langle \eta_{i}(t) \eta_{j}(t') \right\rangle = \delta_{ij}\delta(t-t')$.
The deterministic part $f_{i}$ and the noise prefactor $g_{ij}$ can be written as~\cite{Lau2007}:
\begin{align}
f_{i} = -\Gamma_{ij}\frac{\partial U}{\partial x_{j}}
+ \underbrace{ \frac{1}{2}\frac{\partial}{\partial x_{j}}(g_{ik}g_{jk})
	- a\frac{\partial g_{ij}}{\partial x_{k}}g_{kj}}_{\text{Spurious drift }\nu_{i}} 
\quad\text{and}\quad			      
g_{ik}g_{jk} = 2D_{ij} \, . \label{eq:d-fg}
\end{align}
Here, $U(\{x_{i}\})$ is the potential energy, and $a\in[0,1]$ is the time discretisation parameter as before. 
The second and the third terms in Equation (\ref{eq:d-fg}) constitute the spurious drift $\nu_{i}$, 
whose presence ensures the Boltzmann distribution in the steady state: $P(\{x_{i}\},t\rightarrow\infty)\propto \e^{-U(\{x_{i}\})/T}$. In~(\ref{eq:d-fg})
$\Gamma_{ij}$ and $D_{ij}$ are the mobility and the diffusion matrix, respectively, which must \mbox{satisfy FDT:}
\begin{equation}
D_{ij}(\{x_{i}\}) = \Gamma_{ij}(\{x_{i}\})T \, .
\end{equation}
Onsager symmetry requires $\Gamma_{ij}$ and $D_{ij}$ to be symmetric with respect to $i\leftrightarrow j$ 
and semi-positive definite
(to check this, one can insist $-\left< dU/dt\right>$ to be semi-positive definite). 
Hence, $g_{ij}$ can also be chosen to be symmetric without loss of~generality.

\subsubsection{Conversion from Stratonovich to It\^{o} Integral}\label{sec:elsen_SI}

Now suppose we discretise the time $t$ into $t_{n}=t_{0}+n\Delta t$, where $n=0,1,\dots N$. 
The trajectories $\{x_{i}(t)|t\in[t_{0},t_{N}]\}$ then become $\{x_{i}^{n}|n=0,1,\dots N\}$, and~the discretised Langevin equation reads:
\begin{align}
\Delta x_{i}^{n} &= x_{i}^{n+1}-x_{i}^{n} = f_{i}(\{x_{i}^{n+a}\})\Delta t
+ g_{ij}(\{x_{i}^{n+a}\})\xi_{j}^{n}\sqrt{\Delta t} \\
&= f_{i}(\{x_{i}^{n}\})\Delta t
+ g_{ij}(\{x_{i}^{n}\})\xi_{j}^{n}\sqrt{\Delta t}
+ a\frac{\partial g_{ij}(\{x_{i}^{n}\})}{\partial x_{k}}g_{kl}(\{x_{i}^{n}\})\xi_{j}^{n}\xi_{l}^{n}\Delta t
+ \mathcal{O}(\Delta t^{3/2}) \, ,
\end{align}
where $\{\xi_{i}^{n}\}$ is a set of independent Gaussian random variables with zero mean, $\left\langle \xi_{i}^{n}\right\rangle = 0$,
and Kronecker delta-correlations, $\left\langle \xi_{i}^{m}\xi_{j}^{n}\right\rangle =\delta_{ij}\delta_{mn}$.

Of particular interest below are the It\^{o} ($a = 0$) and Stratonovich ($a=1/2$) discretisations. Let usconsider the following two integrals
\begin{align}
\mathcal{I}_{ij}^{S} & = \int_{t_{0}}^{t_{N}}h_{i}(\{x_{i}\})\eta_{j}(t)\,\textrm{d}t 
:= \sum_{n=0}^{N}h_{i}(\{x_{i}^{n+\frac{1}{2}}\})\xi_{j}^{n}\sqrt{\Delta t} \, , \label{eq:d-Strato}\\
\mathcal{I}_{ij}^{I} & = \int_{t_{0}}^{t_{N}}h_{i}(\{x_{i}\})\cdot\eta_{j}(t)\,\textrm{d}t 
:= \sum_{n=0}^{N}h_{i}(\{x_{i}^{n}\})\xi_{j}^{n}\sqrt{\Delta t} \, . \label{eq:d-Ito}
\end{align}
Here, $h_{i}$ is a general function of $\{x_{i}(t)\}$. 
Note that $\xi_{j}^{n}\sqrt{\Delta t}$ on the right-hand side of (\ref{eq:d-Strato},\ref{eq:d-Ito}) is also called the Wiener process $\int_{t_{n}}^{t_{n}+\Delta t}\eta(t)\,\textrm{d}t$. In~$\mathcal{I}_{ij}^{S}$, the~Stratonovich integral, $h_{i}$ is evaluated at the mid-points $\{x_{i}^{n+\frac{1}{2}}\}$, whereas in the It\^{o} integral,
$\mathcal{I}_{ij}^{I}$, $h_{i}$ is evaluated at the start-points $\{x_{i}^{n}\}$ of each time increment. 
To connect the two integrals, we expand $x_{i}^{n+\frac{1}{2}}=x_{i}^{n}+\frac{1}{2}\Delta x_{i}^{n}$ in Equation (\ref{eq:d-Strato}) to give:
\begin{align}
\mathcal{I}_{ij}^{S} &= \sum_{n=0}^{N}\left[h_{i}(\{x_{i}^{n}\})\xi_{j}^{n}\sqrt{\Delta t}
+ \frac{1}{2}\frac{\partial h_{i}(\{x_{i}^{n}\})}{\partial x_{k}}g_{kl}(\{x_{i}^{n}\})\xi_{l}^{n}\xi_{j}^{n}\Delta t\right]
+ \mathcal{O}(\Delta t^{3/2}) \, .
\end{align}
Finally we can approximate $\xi_{l}^{n}\xi_{j}^{n}\simeq\delta_{lj}$ (which is valid in the limit $\Delta t\rightarrow0$~\cite{Lecomte2017}) \mbox{to obtain}
\begin{align}
\mathcal{I}_{ij}^{S} & =\mathcal{I}_{ij}^{I}+\mathcal{I}_{ij}^{S\rightarrow I} \, ,
\end{align}
where the conversion term $\mathcal{I}_{ij}^{S\rightarrow I}$ is just the (noiseless) Riemann integral
\begin{equation}\label{eq:Elsen-Ito-Strato}
\mathcal{I}_{ij}^{S\rightarrow I}=\frac{1}{2}\int_{t_{0}}^{t_{N}}\frac{\partial h_{i}(\{x_{i}\})}{\partial x_{k}}g_{kj}(\{x_{i}\})\,\textrm{d}t \, .
\end{equation}

\subsubsection{Dynamical~Action}\label{sec:Elsen-d>1_Action}

Following a similar derivation for the case $d=1$ given above, the~path probability for some discretised trajectory $\{x_{i}^{n}|n=0,1,\dots N\}$ is given by
$\mathcal{P}\{x_{i}^{n}\} \propto \e^{-\mathcal{A}\{x_{i}^{n}\}}$,
where the action is~\cite{Lau2007}
\begin{align}
\mathcal{A}\{x_{i}^{n}\} &= \sum_{n=0}^{N-1}\Delta t
\Bigg\{\frac{1}{4}\left(\frac{x_{i}^{n+1}-x_{i}^{n}}{\Delta t}-f_{i}+ag_{ik}\frac{\partial g_{lk}}{\partial x_{l}}\right)
D_{ij}^{-1}\left(\frac{x_{j}^{n+1}-x_{j}^{n}}{\Delta t}-f_{j}+ag_{jk}\frac{\partial g_{lk}}{\partial x_{l}}\right) \nonumber \\
&+ a\frac{\partial f_{i}}{\partial x_{i}}
+ \frac{a^{2}}{2} 
\left(\frac{\partial g_{ik}}{\partial x_{j}}\frac{\partial g_{jk}}{\partial x_{i}}-\frac{\partial g_{ik}}{\partial x_{i}}\frac{\partial g_{jk}}{\partial x_{j}}\right)
\Bigg\} \, ,
\end{align}
where $f_{i}$, $g_{ij}$, $D_{ij}$, and their derivatives are evaluated at $\{x_{i}^{n+a}\}$, and~$D^{-1}$ is the inverse matrix of $D$, with~matrix elements $D^{-1}_{ij}$. 
The transition probability from $\{x_{i}^{0}\}$ at time $t_{0}$ to $\{x_{i}^{N}\}$ at time $t_{N}$ can then be written as a path integral
\begin{equation}
P(\{x_{i}^{N}\}|\{x_{i}^{0}\})=\int\left(\prod_{n=1}^{N-1}\prod_{i}\frac{\textrm{d}x_{i}^{n}}{\det(g_{ij}^{n+a})\sqrt{2\pi\Delta t}}\right)\e^{-\mathcal{A}\{x_{i}^{n}\}}\rightarrow\int\prod_{i}\mathcal{D}x_{i}(t)\,\e^{-\mathcal{A}[\{x_{i}(t)\}]} \, ,
\label{eq:multipath}
\end{equation}
in the limit of $\Delta t\rightarrow0$. 
For future reference, we shall also write:
\begin{equation}
\mathcal{A}\{x_{i}^{n}\} = \sum_{n=0}^{N-1}\Delta t
\left\{ \frac{1}{4}\left(\frac{x_{i}^{n+1}-x_{i}^{n}}{\Delta t}+\Gamma_{ik}\frac{\partial U}{\partial x_{k}}\right)
D_{ij}^{-1}\left(\frac{x_{j}^{n+1}-x_{j}^{n}}{\Delta t}+\Gamma_{jk}\frac{\partial U}{\partial x_{k}}\right)\right\} 
+ \mathcal{A}_{\text{conv}} \, ,
\end{equation}
where $\mathcal{A}_{\text{conv}}$ contains all terms which depend on $a$ explicitly. 
For instance, for~additive noise, where $g_{ij}$, $\Gamma_{ij}$, and $D_{ij}$ are constant, the~$a$-explicit term is
\begin{equation}\label{eq:Elsen-Aconv}
\mathcal{A}_{\text{conv}} = a \sum_{n=0}^{N-1}\Delta t\frac{\partial f_i}{\partial x_i} \, .
\end{equation}

As already described for $d=1$ in Section~\ref{sec:Elsen-EP}, when calculating the EPR, the~preferred choice for the time discretisation is $a=1/2$ (Stratonovich) so that the pre-exponential product in Equation (\ref{eq:multipath}) is the same for any forward and backward pair of paths. With~this choice of $a=1/2$ (only),  $\mathcal{A}_{\text{conv}}$ is identical for the pair and therefore cancels when the difference of their actions is taken to give the~EPR.

\section{Scalar Active Field Theories with Additive~Noise} \label{sec:additive}
We now turn our focus to field-theoretical models. These require discretisation in space as well as time. We will see that the analysis of time-reversibility for fluctuating hydrodynamics brings additional difficulties with respect to finite dimensional systems. In~what follows, we show that these difficulties can be resolved by carefully choosing the spatial discretisation scheme, as~well as the temporal~one.

Throughout this section, we address the fluctuating
hydrodynamics for a single conserved scalar field, governed by diffusive (Model B-like) dynamics. This describes a system that undergoes phase-separation. We allow for activity but~insist that the steady-state EPR must vanish when active terms are switched off.  The~various considerations set out here generalize in varying degrees to more complex models of the kinds mentioned in the~Introduction.

The dynamics of a diffusive conserved scalar order parameter $\phi(\bfr,t)$ is governed by
\begin{equation}
\label{eq:general_AMB}
\dot \phi = -\nabla \cdot (\bfJ_d+ {\bm \Lambda}) \, ,
\end{equation}
where ${\bf J}_d$ is a deterministic current and ${\bm \Lambda}$
a spatio-temporal Gaussian white noise  current satisfying
\begin{equation} \label{eq:AMB-noise}
\avg{ \Lambda_\alpha (\bfr, t) \Lambda_\beta (\bfr',t') } = 
2T\Gamma\,\delta_{\alpha\beta} \delta (\bfr-\bfr') \delta (t-t') \, .
\end{equation}
Here, $T$ is the temperature and $\Gamma$ is the collective mobility. In~principle, $\Gamma = \Gamma[\phi]$, but we now take it to be constant so that the noise is additive~\cite{Gardiner,Lau2007}. This gives vast technical simplifications that we freely exploit below, with~almost no modification to the physics of interest, namely phase separation.
For passive systems en route to equilibrium, the~deterministic part of the
current takes the form
\begin{equation}
\bfJ_d \equiv -\Gamma\nabla \mu \, , \quad  \mu=\mu_\E \equiv \f{ \delta \cF [\phi] }{ \delta \phi } \, .
\end{equation}
This is Model
B~\cite{hohenberg1977theory,Chaikin}.  The~chemical potential $\mu_\E$
derives from a free energy $\cF[\phi]$, which is conveniently chosen of
the $\phi^4$-type
\begin{eqnarray}
\label{eq:AMB}  \cF[\phi] =\int
\brt{ f (\phi)+ \f{\kappa(\phi)}{2} | \nabla \phi |^2 } \D \bfr ,
\qquad
f(\phi) = \f{ a_2 \phi^2 }{2} + \f{ a_4 \phi^4 }{4},
\end{eqnarray}
with $a_4$ and $\kappa(\phi)$ strictly positive. Phase separation then arises, at~mean-field level, whenever $a_2<0$. 

Extensions of Model B have recently played a crucial role in understanding phase separation in active systems. In~the simplest setting~\cite{Wittkowski2014,Nardini2018,thomsen2021periodic,fausti2021}, these theories only retain the evolution of the density field $\phi$, while hydrodynamic~\cite{tiribocchi2015active,singh2019hydrodynamically} or polar~\cite{Tjhung2012,Markovich2019} fields can be added if the phenomenology requires. The~top-down construction of these field theories, via conservation laws and symmetry arguments, closely retraces the path leading to Model B for passive phase separation~\cite{CatesLesHouches}. However, locally broken time-reversal symmetry implies that new non-linear terms are allowed. The~ensuing minimal theory, Active Model B+~\cite{Nardini2017,Nardini2018}, includes all terms that break detailed balance up to order $\mathcal{O}(\nabla^4\phi^2)$ in a gradient expansion of the dynamics of $\dot\phi$~\cite{Nardini2017,Nardini2018}. It is defined by replacing $ {\bf J}$ in Equation (\ref{eq:general_AMB}) by
\begin{align}\label{eq:AMB+}
{\bf J}_d =-\Gamma\left[\nabla \mu  - \zeta (\nabla^2\phi)\nabla\phi\right] \, , \qquad
\mu= \mu_\E +\mu_\A \, ,\qquad \mu_\A=\lambda|\nabla\phi|^2 \, ,
\end{align}
which contains two activity parameters, $\lambda$ and $\zeta$, which are independent in more than one dimension. Model B is recovered at vanishing activity
($\lambda=\zeta=0$)~\cite{hohenberg1977theory}. Note that we retain constant noise amplitude; such noise need not be thermal in origin in an active system, although~it can be in some interesting near-equilibrium cases as will be addressed in Section~\ref{sec:Etienne-Tomer}. This model could be further complemented by a coloured noise, a~feature that has been recently considered~\cite{Paoluzzi2016,maggi2021critical}. Note also that the decomposition of $\mu$ into its equilibrium and nonequilibrium parts is not unique. Since the only defining property of $\mu_\A$ is that it does not derive from a free energy, an~arbitrary equilibrium contribution can be moved into it from $\mu_\E$.  For~simplicity, we set $\Gamma=1$ without loss of generality and~also set $\zeta=0$.  In~addition, we will now choose the (positive) square gradient coefficient to be a constant, $\kappa(\phi) = \kappa$, following~\cite{hohenberg1977theory,Chaikin}. This simplified model was introduced in~\cite{Wittkowski2014} and is known in the literature as Active Model B (AMB); as just described, it is a special case of AMB+ but~sufficient for our present~purposes. 

In analogy with the finite-dimensional case discussed in Section~\ref{sec:Elsen-d>1_Action}, the~action of AMB can be written as
\begin{equation}\label{eq:intro-S} 
\al \cA[\phi] &= 
-\f{1}{4T} \int
\D \bfr \textrm{d}t 
\,
( \dot \phi + \nabla \cdot \bfJ_d) \nabla^{-2} (\dot
\phi + \nabla \cdot \bfJ_d) 
+ \cA_{\textrm{conv}} \, , 
\all
\end{equation}
where $\cA_{\textrm{conv}}$ depends on the scheme employed for the time-discretisation. (Note that the inverse Laplacian in Equation (\ref{eq:intro-S}) is well defined as a Coulomb integral in either an infinite or periodic domain.)
At first sight, it is straightforward to generalise the expression for $\cA_{\textrm{conv}}$ that was given for finite-dimensional systems in Equation (\ref{eq:Elsen-Aconv}) as
\begin{equation}\label{eq:A_conv-1}
\cA_{\textrm{conv}}
=
-a \int \D \bfr \,\D s\, \frac{\delta \nabla\cdot \bfJ_d(\bfr)}{\delta \phi(\bfr)}\,,
\end{equation}
where $s \in [0,t]$, here and below, is a time variable. Importantly, however, no mathematical sense can be given  to Equation (\ref{eq:A_conv-1}) without an explicit choice of {\em spatial} discretisation. Indeed, if~we try to retain continuous spatial variables, from~Equations (\ref{eq:AMB+}) and (\ref{eq:A_conv-1}), we obtain
\begin{eqnarray}\label{eq:A_conv}
\nonumber \cA_{\textrm{conv}}&=&
a\int \D \bfr \,\D s\,\frac{\delta }{\delta \phi(\bfr)} \left[
\nabla^2 
\left(
\mu_E(\bfr)+
\lambda |\nabla \phi(\bfr)|^2
\right) \right] \\
&=&
a\int \D \bfr\, \D s\, \left[
f''(\phi(\bfr))  \nabla^2 \delta(0)
- \kappa \nabla^4 \delta(0)
- 2 \lambda
\nabla\cdot\left\{
\left(\nabla \phi(\bfr)\right)
\nabla^2 \delta(0)
\right\} \right]\,.
\end{eqnarray}

Here, the presence of $\delta(0)$ (the Dirac delta evaluated at zero argument) does not allow a continuum interpretation even in the distributional sense. The~problem arises from the fact that Equation (\ref{eq:A_conv-1}) contains a functional derivative at point $\bfr$ of a function ($\nabla\cdot \bfJ_d$) evaluated at the {\em same} spatial location $\bfr$. As~we shall see in Section~\ref{sub:spatial-discretisation}, a~proper interpretation can be given only after discretising the dynamics in space. We will then find that $\cA_{\textrm{conv}}$ not only diverges as the continuum limit is taken (resulting in the $\delta(0)$ terms), but~that it depends on the spatial discretisation scheme~used. 

\subsection{Informatic Entropy~Production}
It is straightforward to notice that
$\cA_{\textrm{conv}}$ 
is symmetric in time; thus, although~it reweights paths in a configuration-dependent manner, it does not contribute to the steady-state IEPR~\cite{Nardini2017}, which reads
\begin{equation}\label{eq:mec1a} 
\dot{\cS} = 
\left<\lim_{t\to\infty} \f{1}{t} \log \f{ \cP[\{{\bf J}\}_0^{t}] }{
	\cP^\R[\{{\bf J}\}_0^{t}] } \right>
=
-\lim_{t\to\infty} \f{1}{T t}
\int \D \bfr
\int_0^{t}  
\left< \mu_\A \dot \phi \right>  \D s \, , 
\end{equation} 
where the integral over time is performed within the Stratonovich scheme and the average is taken with respect to noise realizations.
For active systems, $\dot{\cS}\ge 0$, with~equality only if, at~the coarse grained scale of the field $\phi(\bfr,t)$, the~emergent dynamics is reversible. It is perfectly possible, in principle~\cite{egolf2000equilibrium}, that reversible dynamics do emerge after coarse graining even though the microscopic processes powering the dynamics of  $\phi$ are very irreversible. However, the~generic case in active matter is, of course, to have irreversible dynamics at the mesoscopic scale described by $\phi(\bfr,t)$, and~hence, have positive IEPR in Equation (\ref{eq:mec1a}).

Recall that in contrast with the case of a forced thermal particle considered in \mbox{Equation~(\ref{eq:1D-entropy})}, but~ just as for the single active particle considered in  Equation~(\ref{eq:1D-driven}), the~{\em informatic} entropy production rate $\dot{\cS}$ given by Equation (\ref{eq:mec1a}) cannot be interpreted as the ratio between the heat produced and the temperature. There are several reasons for this. Firstly, in~a general active setting, even the passive-looking terms in the model (those entering $\mu_\E$) need have no connection with interparticle forces: like the active terms, they could emerge from purely behavioural rules among swimming microorganisms, say. Thus, there is no first law, and~no direct connection with heat. Second, even in a system where these connections can be made and a first law established, to~capture the full heat production of the system, one must consider all microscopic degrees of freedom, not just the coarse-grained fields. However, for~systems whose activity can be viewed as a small departure from thermal equilibrium, there is a middle path in which one can embed an active field theory within a larger model whose thermodynamics is consistent at the level of the degrees of freedom actually retained. This approach was developed in~\cite{Markovich2021} and~will be reviewed in Section~\ref{sec:Etienne-Tomer}. 

Meanwhile, as~explored in~\cite{Nardini2017,Nardini2018,Borthne2020,FodorJackCates2021}, the IEPR has emerged as a useful tool for quantifying the extent to which the behaviour of active complex fluids {\em at hydrodynamic level } (as described by $\phi$ and/or additional order parameters such as fluid velocity, nematic or polar order, etc.) is irreversible. We give an example of such calculations, which can only be performed numerically and therefore requires further consideration of discretisation, in~Section~\ref{sec:t-discr} below.

\subsection{Spatial~Discretisation}\label{sub:spatial-discretisation}
We now discuss spatial discretisation strategies for AMB. The~reason is two-fold. First, as~we have seen above, we are unable to give a precise mathematical meaning to the action $\cA[\phi]$ of a fluctuating hydrodynamic theory working directly at the continuum level; it is natural to expect, and~we confirm this here, that the issue can be solved by discretising the dynamics in space. 
Second, to~numerically integrate any field theory, it is necessary to employ some form of spatial discretisation. A~desirable feature of the discretisation used, which becomes crucial if one is interested in measuring $\dot{\cS}$, is that the ensuing discrete system respects time-reversal symmetry if the field theory one intends to approximate does. We thus describe here how to perform spatial discretisation of AMB such that detailed balance is always recovered in the equilibrium limit for AMB ($\lambda \to 0$). 
For simplicity, we focus on the one-dimensional AMB of finite width $L$ such that $x\in[0,L]$ with periodic boundary conditions; extending these results to higher dimensions is straightforward. (Note also that in one dimension, the~$\zeta$ and $\lambda$ nonlinearities in Equation (\ref{eq:AMB+}) are not independent, so we include AMB+ up to the parameter shift $\lambda\to \lambda-\zeta/2$.)

We discretise $x$ into $N$ lattice points with equal lattice spacing $\Delta x $ so that $N\Delta x =L$, and~the density field as $\phi(x,t)\rightarrow (\phi_{1},...,\phi_N)$; $\phi_{i}(t)$ is the value of $\phi$ at $x=i\Delta x $, where $i=1,2,\ldots N$. Representing the discrete gradient and Laplacian operators as
\begin{align}\label{eq:grad-general}
\nabla \psi_i = \sum_j A_{ij} \psi_j \, ,
\qquad\qquad
\nabla^2 \psi_i =- \sum_j B_{ij} \psi_j \, ,
\end{align}
the discretised dynamics reads
\begin{equation}\label{eq:AMB-discretised}
\p_t \phi_i = - \sum_j  B_{ij} \mu_j + \sqrt{\f{2T}{\Delta x}} \sum_j  A_{ij} \eta_j \, ,
\end{equation}
with $ \avg{ \eta_i(t) \eta_j (t') } = \delta_{ij} \delta(t-t') $. Given the spatial reflection symmetry of the underlying model ($x\to -x$), a~natural choice is to use midpoint spatial discretisation for the gradient operator, which corresponds to the choice $A_{ij}= (\delta_{i+1,j} - \delta_{i-1,j})/(2\Delta x)$, and hence, $B_{ij}= ( -\delta_{i+2,j}+2\delta_{ij}-\delta_{i-2,j})/(2\Delta x)^2$.

In the passive limit $\lambda=0$, $\mu_i = (1/\Delta x)\p \cF/ \p \phi_i$ so that
\begin{equation}\label{eq:MB-discretised}
\p_t \phi_i = - \frac{1}{\Delta x} \sum_j  B_{ij} \f{ \p \cF }{ \p \phi_j} + \sqrt{\f{2T}{\Delta x}} \sum_j  A_{ij} \eta_j \, .
\end{equation}
Notably, to~ensure that the model respects time-reversibility in the passive limit, we are not free in the choice of the discrete gradient and Laplacian operators. Indeed, Equation (\ref{eq:MB-discretised}) respects detailed balance only if ${\bf A} {\bf A}^{\!\mathsf{T}}={\bf A}^{\!\mathsf{T}}{\bf A}={\bf B}$~\cite{Gardiner,VanKampenBook}, corresponding to  $\nabla^2=\nabla\cdot\nabla$ at the discrete level. 
Happily, the~mid-point spatial discretisation indeed satisfies this condition, and so time is reversible as~required.

A separate discretisation issue is to make sense of $\cA_{\textrm{conv}}$ for AMB, which we found to be divergent if computed directly in the continuum limit. From~Equations (\ref{eq:Elsen-Aconv}) and (\ref{eq:MB-discretised}), we obtain
\begin{equation}\label{eq:Aconv_discrete_AMB}
\cA_{\textrm{conv}}=
- a  \int {\rm d}s \,\sum_i \brt{  
	B_{ii}f''(\phi_i)
	+ \kappa \sum_j 
	B_{ij}^2
	+ 2\lambda \sum_{j,k} B_{ij}A_{ji} A_{jk} \phi_k
} \, .
\end{equation}
As expected from Equation (\ref{eq:A_conv}), these terms are divergent as $\Delta x \to 0$. Interestingly, $\cA_{\textrm{conv}}$  not only depends on the choice of the time-discretisation encoded in $a\in[0,1]$ but~also on the choice of the spatial discretisation encoded in the matrices ${\bf A}$ and ${\bf B}$. Still, with~the Stratonovich choice ($a=1/2$), we have that $ \cA_{\textrm{conv}} - \cA_{\textrm{conv}}^\R = 0$. This shows that, even for active fields, $\cA_{\textrm{conv}}$ does not contribute to the IEPR, which we consider~next.

\subsection{Computing the~IEPR}\label{sec:t-discr}
Evaluating the informatic entropy production rate $\dot{\cS}$ in numerical simulations of fluctuating hydrodynamics exposes a subtlety which is once again related to the precise spatial discretisation used. When simulating the dynamics numerically, it is often preferable to employ It\^{o}'s prescription,
so that the update at a given timestep depends only on prior data (thus avoiding use of predictor-corrector or other iterative procedures). However, for~reasons given in Section~\ref{sec:Elsen-EP} above, $\dot{\cS}$ is reliably accessible only within the Stratonovich framework. Following standard stochastic calculus rules as recalled in Section~\ref{sec:elsen_SI} for finite-dimensional systems, one might be tempted to transform the Stratonovich integral defining $\dot{\cS}$ into an It\^{o} integral that in turn can be computed using trajectories obtained directly from integrating the It\^{o}-discretised time dynamics. Subtleties, however, arise when pursuing this path for stochastic PDEs, which can be fully clarified only by also discretising the spatial dynamics as we do~here. 

We again consider the case of AMB, for~which the IEPR is given by Equation (\ref{eq:mec1a}). Working directly at the continuum level, let us first try to transform the Stratonovich integral appearing in Equation (\ref{eq:mec1a}) into an It\^{o} integral by generalising to the infinite dimensional case the conversion term that we have given in Equation (\ref{eq:Elsen-Ito-Strato}) for finite dimensions. We obtain
\begin{equation}\label{eq:ISI-continuous}
\dot{\cS} = -\lim_{t\to\infty} 
\left\{
\f{1}{T t}
\int \D \bfr
\int_0^{t}  
\left<\mu_\A \cdot \dot \phi \right>   \D s +\int \D \bfr\, \left\langle \mathcal{I}_{\textrm{S}\to \textrm{I}}(\bfr,\bfr)\right\rangle
\right\} \, ,
\end{equation} 
where
\begin{align}\label{eq:Ito_Strat_S_cont}
\nonumber \mathcal{I}_{\textrm{S}\to \textrm{I}} (\bfr_1,\bfr_2)
&= 
\frac{1}{t}\int_0^{t}  \D s\,\nabla_{\bfr_2}^\alpha\cdot \frac{\delta \nabla_{\bfr_1}^\alpha \mu_\A(\bfr_1)}{\delta \phi(\bfr_2)} \\
&=
-\frac{2\lambda}{t} 
\int_0^{t}  \D s\,
\nabla_{\bfr_2}^\alpha\nabla_{ \bfr_2}^\beta \big[\left(\nabla_{\bfr_2}^\beta \phi(\bfr_2)\right) \nabla_{\bfr_1}^\alpha\delta(\br_1-\br_2)\big]\,,
\end{align}
in which $\nabla_{\bfr_{\{1,2\}}}$ denotes the gradient operator with respect to $\bfr_{\{1,2\}}$, and~$\alpha,\beta$ are spatial coordinates.
Given that the correction to $\dot{\cS}$ is given by an integral over space of $ \mathcal{I}_{\textrm{S}\to \textrm{I}} (\bfr,\bfr)$, and~that the latter is a divergence, one might speculate that there is no correction due to the Stratonovich to It\^{o} transformation (at least for periodic boundary conditions). However, taking $\br_1=\br_2$ in Equation  (\ref{eq:Ito_Strat_S_cont}), as~required to evaluate Equation (\ref{eq:ISI-continuous}), produces an undefined $\delta(0)$ divergence.

We, therefore, consider the entropy production rate of the fully discretised dynamics (\ref{eq:AMB-discretised}) and perform the same transformation from Stratonovich to It\^{o} integral:
\begin{align}
\dot{S_d} 
= 
-\lim_{t\to\infty} \f{\Delta x}{T t} \sum_i\int_0^{t} \left< \mu_{\A,i} \,\dot{\phi}_i \right> \D s=
-\lim_{t\to\infty}\left\{
\sum_i\f{\Delta x}{T t} \int_0^{t} \left< \mu_{\A,i} \cdot\dot{\phi}_i \right> \D s
+\left\langle \mathcal{I}_{\textrm{S}\to \textrm{I}}\right\rangle
\right\} \, ,
\end{align} 
where, using Equations (\ref{eq:Elsen-Ito-Strato}) and (\ref{eq:AMB-discretised}), we have
\begin{equation}\label{eq:AMB-Ito-Strato-discrete}
\mathcal{I}_{\textrm{S}\to \textrm{I}}
=
\sum_{i,j}\frac{B_{ij} }{ t } \int_0^{t} \D s \frac{\p \mu_{\A,i}}{\p\phi_j}\,.
\end{equation}
In the midpoint spatial discretisation, $\mu_{\A,i}$ depends only on $\phi_{i\pm 1}$, while $B_{ij}\neq 0$ only when $j=i,i\pm 2$. In~this case, we thus obtain from Equation (\ref{eq:AMB-Ito-Strato-discrete}) that 
$\mathcal{I}_{\textrm{S}\to \textrm{I}}=0$. 
This is, however, not generic and due to the specific form of the non-equilibrium chemical potential $\mu_A$ of AMB. For~example, suppose we had written the IEPR in the following equivalent form, which includes the reversible part of the chemical potential $\mu_\E$ (whose contribution to $\dot{\cS}$ is a total time derivative that gives zero in the large $t$ limit):
\begin{equation} \label{eq:mec1a-1} 
\dot{\cS} = -\lim_{t\to\infty} \f{1}{T t}
\int \D \bfr
\int_0^{t}  
\left<\mu \dot \phi \right> \D s\,. \end{equation} 
Then the same line of reasoning shows that the Stratonovich to It\^{o} conversion factor does not vanish even within the spatial midpoint discretisation scheme. However, with~either choice of definition for $\dot{\cS}$,
the discrete dynamics, as formulated above, is unambiguous and necessarily leads to the same final result; this has indeed been checked numerically for AMB~\cite{Nardini2017}.

Let us now revisit the computation that we attempted at the continuum level with Equations~(\ref{eq:ISI-continuous}--\ref{eq:Ito_Strat_S_cont}). If~we employ the following definition of the operator $\nabla_{\bfr} \frac{\delta }{\delta \phi(\bfr)}$ acting on  arbitrary functions $g$ of $\phi$ and its derivatives:
\qq\label{eq:def-op}
\nabla_{\bfr} \frac{\delta g(\bfr)}{\delta \phi(\bfr)}
\equiv
\left.\nabla_{\bfr_2} \frac{\delta g(\bfr_1)}{\delta \phi(\bfr_2)}\right|_{\bfr_1=\bfr_2=\bfr}
=
\lim_{\Delta x\to0}
\sum_{j} \frac{A_{ij}}{\Delta x} \frac{\p g_i}{\p\phi_j}\,,
\qqq
{we }  
obtain
\qq
\int \D \bfr   \mathcal{I}_{\textrm{S}\to \textrm{I}}(\bfr,\bfr)
=
\lim_{\Delta x\to 0}
\mathcal{I}_{\textrm{S}\to \textrm{I}}\,,
\qqq
where $\mathcal{I}_{\textrm{S}\to \textrm{I}}$ obeys Equation (\ref{eq:AMB-Ito-Strato-discrete}) and depends on the discretisation scale $\Delta x$. 
It should be observed, however, that Equation (\ref{eq:def-op}) remains only a formal relation because~the right-hand side can be a divergent quantity. This underlines the fact that to avoid all conceptual ambiguities, we should work with a finite discretisation~length.

\begin{figure}[h]
	\center
	\includegraphics[width=1\linewidth]{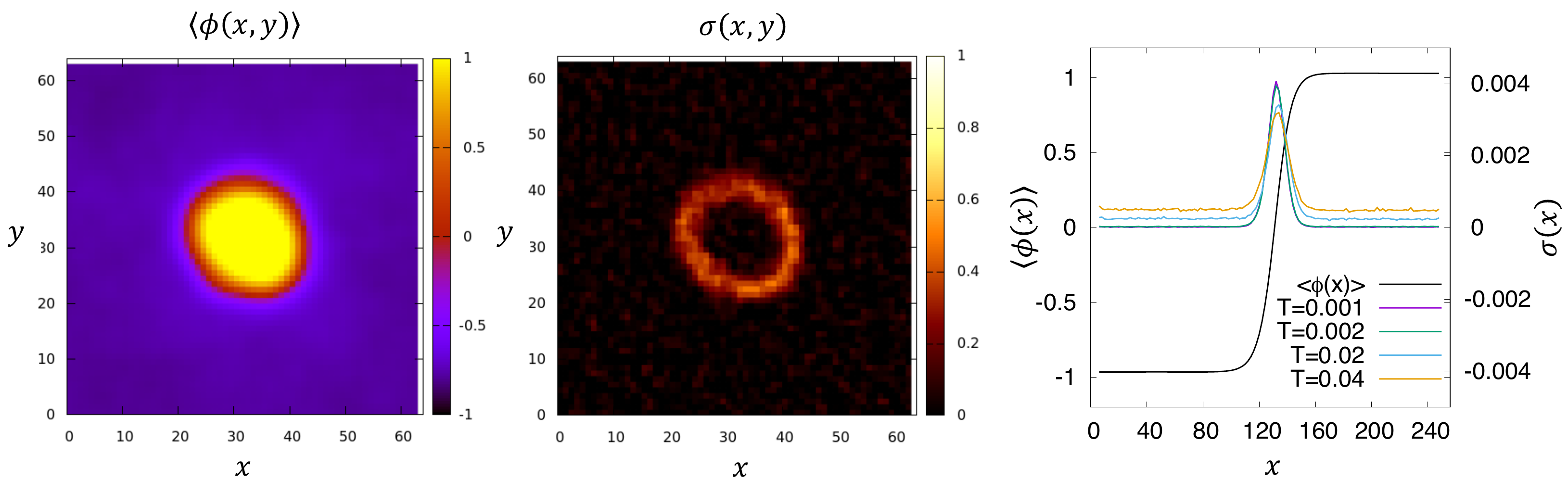}
	\caption{\label{fig:sigmarealspace}
		{Adapted from}    
		~\cite{Nardini2017}. ({\bf Left}) Density map of a fluctuating phase-separated droplet in two-dimensional AMB. ({\bf Center}) Local contribution to the informatic entropy production \mbox{$\sigma({\bf r}) = -\lim_{t\to\infty} \f{1}{T t}
			\int_0^{t}  
			\langle\mu_\A\dot \phi\rangle(\bfr,s) \, \D s$} showing a strong contribution at the interfaces. ({\bf Right}) Density and entropy production for a 1D system comprising a single domain wall for~various temperatures $T\ll a_2^2/4a_4$. The~entropy production is strongly inhomogeneous, attaining a finite value as $T\to0$ at the interface between dense and dilute regions and converging to zero in the bulk in this limit. Values of the parameters used are: $a_2=-0.125$, $a_4=0.125$, $\kappa=8$, $\lambda=2$, $\Delta x = 1$, and~$\Delta t = 0.01$.
	}
\end{figure}

The above analysis shows that, when computing numerically the entropy production rate for field theories, care must be taken with not only the temporal but also the spatial discretisation employed. Using the methodology reviewed here, 
$\dot \cS$ was computed numerically within AMB in~\cite{Nardini2017}. Since
this quantity is written as a spatial integral, it is natural in the steady state to associate the first integrand in Equation (\ref{eq:ISI-continuous}), 
$\sigma(\bfr)  = -\lim_{t\to\infty} T^{-1}t^{-1}\int_0^t\langle\mu_\A\dot\phi\rangle(\bfr,s)\,\D s$, with~a {\em local IEPR density}. 
When the steady-state system is phase separated, it was shown that for a small $T$, this density is concentrated at the interfaces between the liquid and vapour phases; see Figure \ref{fig:sigmarealspace}, where it scales as $T^0$. Away from interfaces in the bulk of each fluid, it instead scales as $T^1$. Notably, in~active field theories that show deterministic currents in the steady state (such as the uniformly aligned state of a polar active liquid crystal~\cite{Borthne2020}), the~IEPR density diverges as $T^{-1}$. Observing such scalings numerically can give insight into how and where in the system the active dynamics breaks time reversal symmetry~\cite{FodorJackCates2021}.


\unskip


\section{Thermodynamics of Active Field~Theories}\label{sec:Etienne-Tomer}

In this section, we review what happens when active field theories are minimally coupled to chemical degrees of freedom~\cite{Markovich2021}. The~latter can describe the energy flows underlying activity so long as the active motion itself results from locally weak departures from the thermal equilibrium. This allows the recreation of a first law. We will show in these extended theories, analogous ambiguities to those encountered in the previous section arise not just when computing the IEPR but even in defining the stochastic dynamics itself. (This is because multiplicative noise arises in the off-diagonal couplings between the two sectors.) As we shall see, these ambiguities are likewise resolved by careful~discretisation.

\begin{figure}[h]
	\centering
	\includegraphics[width=0.5\linewidth]{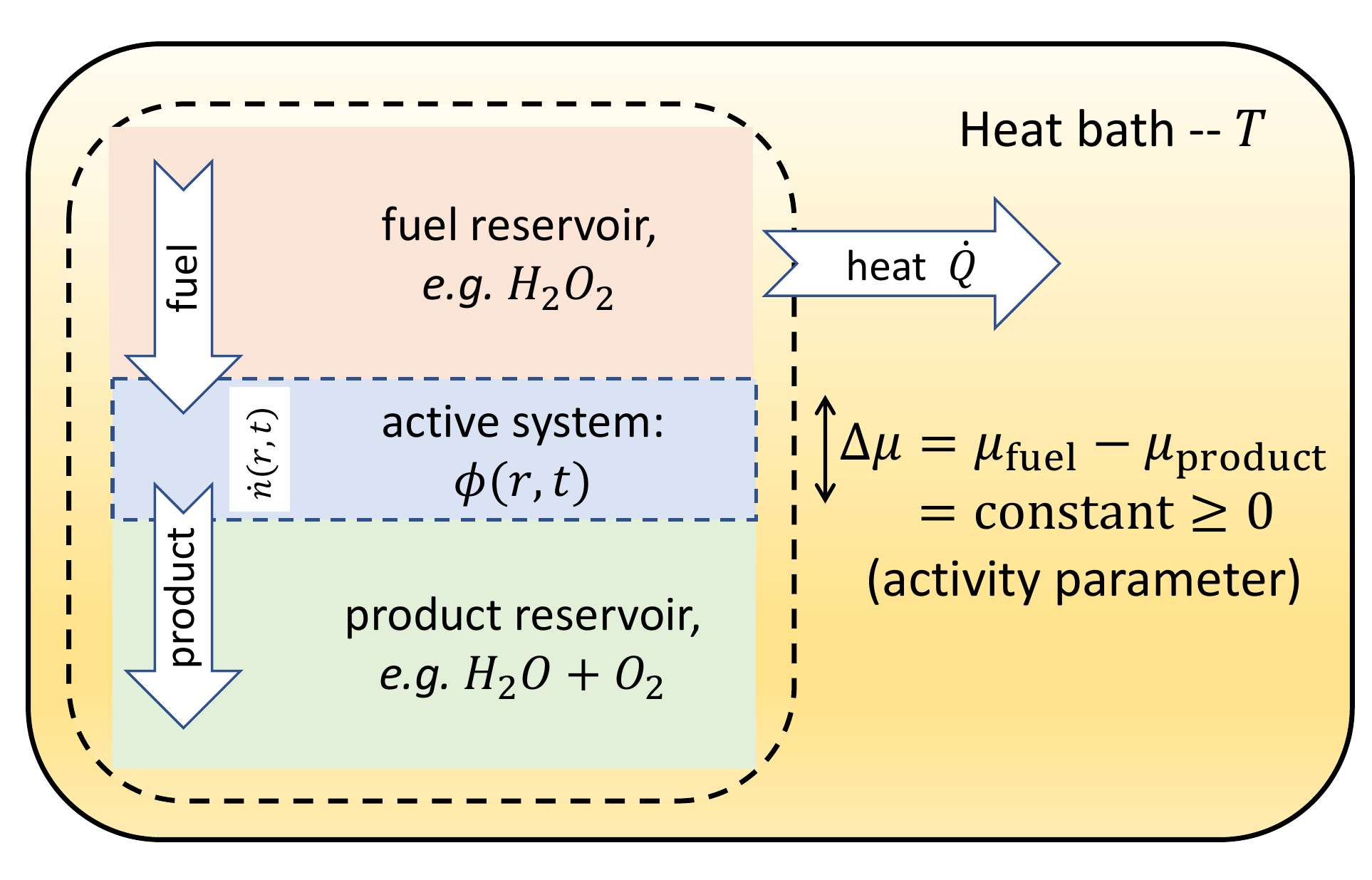}
	\caption{\label{fig1}
		{Schematic representation }  
		of an active system (blue) put in contact with reservoirs of chemical fuel (red) and product (green), which set a constant, homogeneous chemical potential difference $\Delta\mu$ in the active system. Within~our framework, $\Delta\mu$ embodies the driving parameter which controls the nonequilibrium terms in the dynamics Equations (\ref{eq:dyn_phi}) and (\ref{eq:dyn_chem_g}) for the active density field $\phi$ and the rate of fuel consumption $\dot{n}$. The~active system and the chemical reservoirs are surrounded by the thermostat (yellow), which maintains a fixed temperature $T$. The~fluctuations of $\phi$ and $n$ lead to the dissipation of heat ${\cal Q}$ into the thermostat, which quantifies the energetic cost to maintain the whole system away from equilibrium. Note that the physical separation of the reservoirs from the active system, as~illustrated, is conceptually helpful but not necessary: in practice, the fuel, active particles and products can all share the same physical domain. Adapted from~\cite{Markovich2021}.}
\end{figure}

Our interest is in fluctuating hydrodynamic models of complex fluids in which the activity of a conserved scalar field stems from local consumption of chemical fuel. Prototype examples for such active systems are bacterial suspensions~\cite{Yeomans2012, Goldstein2013}, acto-myosin networks~\cite{Fred2010}, and~self-propelling Janus colloids~\cite{Golestanian2007, Bechinger2013, Palacci2013}. At~the continuum level, we therefore address below Active Model B+, as presented in Section~\ref{sec:additive}, which is the leading-order theory of this type.
Activity is assumed to be sustained by connecting the active system to reservoirs of fuel and its products; see Figure~\ref{fig1}. 
Our approach relies on systematically constructing the dynamics of the underlying chemical driving field from that of the active field dynamics based on the force-current relations of Linear Irreversible Thermodynamics (LIT), which obey Onsager reciprocal relations~\cite{Onsager1931}. This physically requires that the activity stems from relatively small departures from the local chemical equilibrium. The~more microscopic the scale of activity or self-propulsion, the~more likely this is to be true: our focus is thus on subcellular systems, or~perhaps Janus colloids, rather than collections of animals~\cite{Toner1995}. 

Importantly, in~some cases, we can construct the extended model (and its discretisation) so that the evolution of the original active fields remains independent of the additional chemical dynamics. This is what we mean by `embedding' the active field theory into a larger model for which thermodynamic consistency and the first law can reappear; we are not {\em changing} the active field theory, just placing it into a more general setting. 
By accounting for the driving mechanism, we find that the rate of heat production for the active system follows from the {\em full} entropy production rate (EPR) measuring the irreversibility of {\em both} the active and driving fields, which can however be evaluated from the fluctuations of active fields only. Importantly, the~heat rate is distinct from the IEPR, $\dot{\cS}[\phi]$, which quantifies the irreversibility of the active field dynamics alone, as~previously described.

As stated above, we will find that the coupling of an active field to its driving mechanism generally results in multiplicative noise~\cite{Gardiner}. It is well known that when dealing with multiplicative (state-dependent) noise, one has to define the specific way in which the noise is evaluated, which affects the time discretisation scheme~\cite{Lau2007} and generally results in spurious drift terms as we discussed for finite dimensional systems, rather than fields, in~Section~\ref{sec:1D}. Moreover, for~the reasons already described in Section~\ref{sec:additive}, we also need to pay careful attention to spatial~discretisation.


\subsection{Onsager Coupling in Two-Dimensional~System} \label{sec:onsager_2d}

Before constructing our thermodynamic active field theory, it is instructive to consider a simple example of a two-particle system in which the single particle dynamics seem to be additive, but~Onsager reciprocal relations result in multiplicative noise due to cross-coupling in the noise~terms.

As a minimal particle-based model for this, let us consider the following dynamics
\begin{equation}\label{eq:ef_dyn}
\begin{aligned}
\dot x &= - \Gamma_x \partial_x V - C(x,y) \partial_y U + T \nu_x + \xi_x \, ,
\\
\dot y &= - \Gamma_y \partial_y V - C(x,y) \partial_x U + T \nu_y + \xi_y \, ,
\end{aligned}
\end{equation}
where $\{\Gamma_x,\Gamma_y\}$ are mobilities, $C$ an arbitrary function of $\{x,y\}$, $T$ the temperature, and~$U$ the potential. Here, $\{\nu_x,\nu_y\}$ are spurious drift terms that will be defined precisely below. The~terms $\{\xi_x, \xi_y\}$ are Gaussian white noises with zero mean and correlations given by
\begin{equation}\label{eq:ef_noise}
\begin{aligned}
\big\langle \xi_x(t) \xi_x(t') \big\rangle &= 2 \Gamma_x T \delta(t-t') \, ,
\\
\big\langle \xi_y(t) \xi_y(t') \big\rangle &= 2 \Gamma_y T \delta(t-t') \, ,
\\
\big\langle \xi_x(t) \xi_y(t') \big\rangle &= 2 C(x,y) T \delta(t-t') \, .
\end{aligned}
\end{equation}
The dynamics in Equations~(\ref{eq:ef_dyn}--\ref{eq:ef_noise}) can be written in a compact form as
\begin{equation}
\begin{aligned}
&\big[ \dot x, \dot y] = - {\mathbb L} \big[\partial_x U, \partial_y U \big] + T \big[\nu_x, \nu_y\big] + \big[\xi_x, \xi_y\big] \, ,
\\
&\big\langle\big[\xi_x, \xi_y\big](t) \big[\xi_x, \xi_y\big]^{\mathsf{T}}(0)\big\rangle = 2 T {\mathbb L} \delta(t) \, ,
\end{aligned}
\end{equation}
where $^\mathsf{T}$ denotes transpose, and~we have introduced the Onsager matrix $\mathbb L$ given by
\begin{equation}
{\mathbb L} =
\begin{bmatrix}
\Gamma_x & C(x,y)
\\
C(x,y) & \Gamma_y
\end{bmatrix}
\, .
\end{equation}
Such a form for linear coupling between the velocities $\{\dot x, \dot y\}$ and the forces $\{-\partial_x U, -\partial_y U\}$ is inspired by the seminal work of Onsager~\cite{Onsager1931}, which demonstrated that $\mathbb L$ must be positive semi-definite (i.e., $\det{\mathbb L}\geq0$) for stability.

Due to the fact that the correlations between $\xi_x$ and $\xi_y$ depend explicitly on $\{x,y\}$ through $C$, one has to specify the time discretisation of Equation~\eqref{eq:ef_dyn}. Changing time discretisation affects the explicit expression of the spurious drift terms $\{\nu_x,\nu_y\}$, which depend on $\{\Gamma_x,\Gamma_y,C\}$ and derivatives of $C$. In~practice, we choose the spurious drift terms at a given time discretisation to ensure that the corresponding Fokker--Planck Equation (FPE) for the probability density $P(x,y)$ reads
\begin{equation}
\begin{aligned}
\dot P &= \partial_x \Big[ \big( \Gamma_x \partial_x U + C(x,y) \partial_y U + \Gamma_x T \partial _x \big) P \Big]
\\
&\quad + \partial_y \Big[ \big( \Gamma_y \partial_y U + C(x,y) \partial_x U + \Gamma_y T \partial _y \big) P \Big]
\\
&\quad + T \Big[ \partial_x \big( C(x,y) \partial_y P \big) + \partial_y \big( C(x,y) \partial_x P \big) \Big] \, .
\end{aligned}
\end{equation}
Then, the~steady-state solution is given by the Boltzmann distribution, $P_{\rm s} \sim \e^{-U/T}$, as~expected for any equilibrium~dynamics.

To compute the EPR, $\dot{\cal S} = \lim_{t\to\infty} ({\cal A}^\R-{\cal A}) / t$, it is convenient to express the dynamic action $\cal A$ (and its time-reversed counterpart ${\cal A}^\R$) associated with dynamics~\eqref{eq:ef_dyn} using the Stratonovich convention (SC). 
Using Equation~\eqref{eq:d-fg}, the~spurious drift terms in SC can then be written as
\begin{equation}\label{eq:dyn_sc}
\begin{aligned}
\nu_x &= {\mathbb M}_{11} \big( \partial_x {\mathbb M}_{11} + \partial_y {\mathbb M}_{21} \big) + {\mathbb M}_{12} \big( \partial_x {\mathbb M}_{12} + \partial_y {\mathbb M}_{22} \big) \, ,
\\
\nu_y &= {\mathbb M}_{21} \big( \partial_x {\mathbb M}_{11} + \partial_y {\mathbb M}_{21} \big) + {\mathbb M}_{22} \big( \partial_x {\mathbb M}_{12} + \partial_y {\mathbb M}_{22} \big) \, ,
\end{aligned}
\end{equation}
where $\mathbb M$ is defined by ${\mathbb M}^2 = \mathbb L$. In~practice, decomposing $\mathbb L$ in terms of the diagonal matrix $\mathbb D$ (with eigenvalues of $\mathbb L$ as entries) and of the projector $\mathbb P$ (constructed from eigenvectors of $\mathbb L$), one obtains ${\mathbb M} = {\mathbb P}^{-1} {\mathbb D}^{1/2} {\mathbb P}$. The~action follows as
\begin{equation}
\begin{aligned}
{\cal A} &= \frac{1}{4T} \int_0^t {\mathbb X} {\mathbb L}^{-1} {\mathbb X}^{\mathsf{T}} \D s +  {\cal A}_{\rm conv} \, ,
\\
{\mathbb X} &= \big[ \dot x, \dot y] + {\mathbb L} \big[\partial_x U, \partial_y U \big] \, .
\end{aligned}
\end{equation}
The term ${\cal A}_{\rm conv}$ is a result of the stochastic time integral in the dynamic action and depends on its interpretation (see~\cite{Lau2007} and Section~\ref{sec:Elsen-d>1_Action}). It is, however, invariant under time reversal and~thus does not contribute to the EPR. Notably, because~we use SC, the spurious drift terms $\{\nu_x,\nu_y\}$ do not appear in the first term of $\cal A$~\cite{Lau2007}. We deduce the \mbox{EPR as}
\begin{align}
\nonumber\dot{\cal S} &= - \frac{1}{2T} \underset{t\to\infty}{\lim} \int_0^t \big[ \dot x, \dot y \big] ( {\mathbb L}{\mathbb L}^{-1} + {\mathbb L}^{-1}{\mathbb L} ) \big[\partial_x U, \partial_y U \big]^\mathsf{T} \D s \\
&= - \frac{1}{T} \underset{t\to\infty}{\lim} \int_0^t \big( \dot x \partial_x U + \dot y \partial_y U \big) \D s \, .
\end{align}
Note that the product in the integrand is written within SC. Then, we can use the standard chain rule $\dot U = \dot x \partial_x U + \dot y \partial_y U$, leading to $\dot{\cal S} = \lim_{t\to\infty} (U(0)-U(t))/(Tt)$, which vanishes provided that $U$ does not change in time. Therefore, we have shown that the dynamics~\eqref{eq:ef_dyn} are associated with vanishing EPR, as~expected at~equilibrium.



\subsection{Spatial Discretisation in Stochastic~Field-Theories}\label{sec:ffpe}

The example above makes it clear that our construction of the underlying driving field using LIT is likely to result in multiplicative noise due to cross-coupling noise terms. Therefore, prior to actually constructing our theory, it is useful to discuss the space-discretisation issue that arises in stochastic field theories with multiplicative noise. This issue is very similar to the one encountered in Section~\ref{sec:additive} for additive noise in the dynamic action of a stochastic field theory, but~here, the problem appears already at the Langevin dynamics. To~present the discretisation issue in the simplest case, we consider the 1D functional diffusion equation for the density $\phi$ of a (thermodynamically) ideal gas with density-dependent diffusivity. In~Appendix~\ref{app:sd_lit}, we provide a more general form of the spurious drift terms within LIT. 



The 1D functional FPE for the density of an ideal gas is~\cite{Solon2015,Gardiner}
\begin{eqnarray}
\label{eq:fpe_diffusion}
\nonumber& &\frac{\partial P[\phi]}{\partial t} = - \int\D x \partial_x \frac{\delta J(\left[\phi(x)\right])}{\delta\phi(x)} \, , \\
& & J([\phi]) = \left[-D(x,[\phi]) \,\partial_x \phi - D(x,[\phi]) \phi \,\partial_x \frac{\delta}{\delta \phi(x)} \right]  P[\phi] \, ,
\end{eqnarray}
which have a steady-state solution $P_{\rm s} \sim \exp(-{\cal F}/T)$, with~${\cal F} = T\int \D x  \left[\phi\ln\phi - \phi\right]$ being the ideal-gas free energy. Here, $D(x,[\phi])$ is a functional of the density field $\phi$, which we take to be purely local so that $D(x,[\phi]) = D(x,\phi(x),\partial_x\phi(x)+...)$. This locality will lead below to strong dependence on the discretisation scale along lines seen already in Section~\ref{sec:additive}.
The corresponding It\^{o}-Langevin equation is~\cite{Solon2015}
\begin{align}
\label{eq:diffusion1}
&\dot\phi(x,t) = -\partial_x J(x,t) \, ,\\ 
\label{eq:diffusion2}
\nonumber&J(x,t) = -D(x,[\phi(x,t)]) \partial_x \phi(x,t) + \phi(x,t)\partial_x \frac{\delta D(x,[\phi(x,t)])}{\delta \phi(x)} \\
&\qquad\,\,\,\,+ \sqrt{2T}{\mathbb M}(x,t) \xi(x,t) \, ,
\end{align}
where $\xi$ is a zero mean Gaussian noise with variance $\langle \xi(x,t)\xi(x',t') \rangle = \delta(x-x')\delta(t-t')$ and ${\mathbb M}^2 = D\phi/T$, such that the fluctuation dissipation theory is obeyed. The~second term in the right hand side of Equation~\eqref{eq:diffusion2} is the spurious drift in the It\^{o} convention, which depends on the noise convention and therefore on the time-discretisation scheme~\cite{Lau2007,Basu2008}.  For~example, in~the Stratonovich convention, this term is changed to $T{\mathbb M}(x,[\phi(x,t)])\partial_x \delta {\mathbb M}(x,[\phi(x,t)]) / \delta \phi(x)$. 

We already see that evaluating the spurious drift above in the continuum description is problematic~\cite{Solon2015,Markovich2021}. The~same issue also arose in Section~\ref{sec:additive}, but~only at the level of the dynamic action. Discretising the dynamics in space solved the issue and revealed the actual meaning of $\partial_x \left[\delta D(x) / \delta \phi(x) \right]$; see Equation~\eqref{eq:def-op}.
Now that we understand the meaning of Equation~\eqref{eq:diffusion2}, and~specifically the spurious drift term, it is straightforward to show how different choices of spatial discretisation result in different spurious drifts. As~a purely mathematical example, consider a system that obeys Equation~\eqref{eq:diffusion1} with $D(x,[\phi]) = \bar D+ \partial_x(\partial_x\phi)^2 $, with~some constant $\bar D$.  The~nonconstant part $D-\bar D$ can be written as either $\partial_x(\partial_x\phi)^2$ or $2(\partial_x\phi)\partial_x^2\phi$, which, after discretisation, become, respectively:%
\begin{equation}\label{eq:Cphi-discretised}
\begin{aligned}
D_i^{(1)} &= \sum_{k,l,m} A_{ik} (A_{kl}\phi_l)(A_{km}\phi_m) \, ,
\\
D_i^{(2)} &= 2 \sum_{k,l,m} (A_{ik} \phi_k) (A_{il} A_{lm}\phi_m) \, .
\end{aligned}
\end{equation}
These of course coincide in the continuum limit, $\Delta x \to 0$. A~priori, one might expect the spurious drift terms to be independent of this choice of implementation, yet we now show that this is not the case. For~$D^{(1)}$, we obtain
\begin{equation}\label{eq:spur_1}
\begin{aligned}
\sum_j A_{ij} \frac{\partial D_i^{(1)}}{\partial\phi_j} &= 2 \sum_{j,k,l} A_{ij} A_{ik} A_{kj} A_{kl} \phi_l
\\
&= - 2 \sum_{j,k,l} (A_{ij} A_{jk}) A_{ik} A_{kl} \phi_l
\\
&= - 2 \sum_{k,l} \big[{\bf A}^2\big]_{ik} A_{ik} A_{kl} \phi_l \, ,
\end{aligned}
\end{equation}
where we have used $A_{ij}=-A_{ji}$. Taking $A_{ij} = (\delta_{i+1,j} - \delta_{i-1,j})/(2\Delta x)$, we deduce $[{\bf A}^2]_{ik} A_{ik}=0$, so that~Equation \eqref{eq:spur_1} is zero. Substituting~Equation \eqref{eq:spur_1} into Equation~\eqref{eq:diffusion2}, we conclude that there is no spurious drift associated with $D^{(1)}$. (However, this no longer holds when considering higher-order schemes for the gradient matrix $\bf A$). Choosing instead $D^{(2)}$, we obtain
\begin{equation}\label{eq:spur_2}
\begin{aligned}
\sum_j A_{ij} \frac{\partial D_i^{(2)}}{\partial\phi_j}
&= 2 \sum_{j,k,l} A_{ij} \big( A_{ij} A_{il} A_{lk} + A_{il} A_{lj} A_{ik} \big) \phi_k
\\
&= - 2 \sum_{j,k,l} \big[ (A_{ij}A_{ji}) (A_{il}A_{lk}) + (A_{il}A_{lj}A_{ji}) A_{ik} \big] \phi_k
\\
&= - 2 \sum_k \big( \big[{\bf A}^2\big]_{ii} \big[{\bf A}^2\big]_{ik} + \big[{\bf A}^3\big]_{ii} A_{ik} \big) \phi_k \, ,
\end{aligned}
\end{equation}
where we used again $A_{ij}=-A_{ji}$. Given that ${\bf A}$ is anti-symmetric, any odd (even) power of ${\bf A}$ is anti-symmetric (symmetric), so that $[{\bf A}^3]_{ii}=0$ and $[{\bf A}^2]_{ii}\neq 0$. Then, Equation~\eqref{eq:spur_2} is always non-zero for any form of the gradient matrix $\bf A$. This simple example of a 1D ideal gas with density-gradient-dependent diffusivity illustrates that the choice of spatial discretisation can drastically affect the form of the spurious drift terms. Although~the chosen form for $D$ is somewhat contrived in this context, we will see that precisely the same discretisation choice will enter our discussion below of spurious drift terms for Active Model B+.

\subsection{Thermodynamics of a Conserved Active Scalar~Field}

We now consider the fluctuating hydrodynamics of a conserved active scalar field. Suitable models can be either obtained from explicit coarse-graining of microscopic dynamics~\cite{Solon2015,Fodor2016,MarLub2021} or~written from symmetry arguments~\cite{Marchetti2013, Marchetti2018}---the~prototypical example of the latter route being Active Model B+, Equation~(\ref{eq:AMB+}). 
The key to embedding such models within a thermodynamic framework is to realize that they omit degrees of freedom (chemical or other), which provide the drive needed to sustain nonequilibrium activity, as~described in Figure~\ref{fig1}~\cite{Markovich2021}.
Therefore, our approach consists of introducing an additional field, associated in this case with chemical reactions that drive the dynamics away from equilibrium. We then identify the nonequilibrium terms in the original dynamics as a coupling to chemical reservoirs following the framework of LIT~\cite{Mazur}. 

The dynamics of a conserved scalar field $\phi$ representing the density of active components for an isotropic material can generally be written as:
\begin{equation}\label{eq:dyn_phi}
\dot\phi = - \nabla\cdot{\bf J} ,
\quad
{\bf J} = - \Gamma \nabla\frac{\delta\cal F}{\delta\phi} + \Delta\mu\,{\bf C} + T \,{\boldsymbol\nu}({\bf C}) + {\boldsymbol\Lambda} \, ,
\end{equation}
where $\cal F$ is the free energy,  $\Gamma$ is the mobility, the~activity term ${\bf C}$ is a vector-valued function of $\phi$ and its gradients, $T$ is the temperature of the surrounding heat bath, and~${\boldsymbol\nu}$ a spurious drift discussed below. %
The driving force for activity is $\Delta\mu$, the~chemical potential difference between fuel and products~\cite{Ramaswamy2017, Ramaswamy2018, Kruse2004}; see Figure~\ref{fig1}. (This is not connected with the chemical potential of the $\phi$ field, as defined in Section~\ref{sec:additive}, and~here denoted $\delta{\cF}/\delta\phi$.) An example of such a reaction is the decomposition of hydrogen peroxide involved in the self-propulsion of Janus colloids~\cite{Golestanian2007, Bechinger2013, Palacci2013}. In~what follows, $n$ is described as a field fluctuating in space and time, while $\Delta\mu$ is kept constant and homogeneous. This would be an appropriate approximation for large fuel/product reservoirs and when the chemical fuel and products diffuse much faster than the active particles within the active system~\cite{Markovich2021}. Note that for Active Model B, we have $\Delta\mu{\bf C} = - \Gamma \lambda\nabla|\nabla\phi|^2$. 

To account for the chemical reactions, we introduce the chemical coordinate $n$, which is (half) the difference between the local number density of product molecules and that of the fuel molecules.
Because the active system is a part of a large nonequilibrium system that relaxes (slowly) towards equilibrium, the~explicit dynamics of $n$ can be deduced from LIT~\cite{Mazur, Basu2008, Prost2017, Julicher2018, Markovich2019b}, in which the thermodynamic fluxes are written as a linear combination of the thermodynamic forces. Identifying ${\bf J}$ and $-\nabla(\delta{\cal F}/\delta\phi)$ as the current and the thermodynamic force associated with $\phi$, respectively, LIT states that (in the absence \mbox{of noise)}
\begin{equation}
\begin{aligned}
\begin{bmatrix}
{\bf J} \\ \;\dot n 
\end{bmatrix}
&= {\mathbb L}
\begin{bmatrix}
\;-\nabla(\delta{\cal F}/\delta\phi)\; \\ \Delta\mu 
\end{bmatrix}
\, ,
\end{aligned}
\end{equation}
where ${\mathbb L}$ is the Onsager matrix. 
It is clear from~Equation \eqref{eq:dyn_phi} that the factor coupling the current $\bf J$ and the force $\Delta\mu$ is directly given by ${\bf C}$ (similarly to what we have seen in Section~\ref{sec:onsager_2d}). Note that, though~LIT states linear relations between forces and currents, the~coupling factor $\bf C$ need not be linear in $\phi$ or its gradients. Accordingly, and~because $\phi$ is even under time-reversal, Onsager reciprocity relations require that the coupling factor between the current $\dot n$ and the force $-\nabla(\delta{\cal F}/\delta\phi)$ is also $\bf C$~\cite{Onsager1931}. The~dynamics of $n$ follows as
\begin{equation}\label{eq:dyn_chem_g}
\dot n = \gamma\Delta\mu - {\bf C}\cdot\nabla\frac{\delta\cal F}{\delta\phi} + T \,\chi({\bf C}) + \xi \, ,
\end{equation}
where $\gamma$ is the chemical mobility, which we take constant in what follows. As~a result of this assumption, the~equation for $\phi$ is autonomous and does not rely on knowing the fluctuations of the chemical field $n$. 

In the above, the~noises $\boldsymbol\Lambda$ and $\xi$ are  Gaussian with zero mean and their correlations are given by
\begin{eqnarray}\label{eq:noise}
& &\big\langle\Lambda_\alpha({\bf r},t)\Lambda_\beta({\bf r}',t')\big\rangle = 2\Gamma T \delta_{\alpha\beta}\delta({\bf r}-{\bf r}')\delta(t-t') \, , \\
& &\big\langle\xi({\bf r},t)\xi({\bf r}',t')\big\rangle = 2\gamma T\delta({\bf r}-{\bf r}')\delta(t-t') \, ,\\
& &\big\langle\Lambda_\alpha({\bf r},t)\xi({\bf r}',t')\big\rangle = 2 T C_\alpha({\bf r},t)\delta({\bf r}-{\bf r}')\delta(t-t') \, .
\end{eqnarray}
The terms $T{\boldsymbol\nu}$ in Equation (\ref{eq:dyn_phi}) and $T{\chi}$ in Equation (\ref{eq:dyn_chem_g}) are direct generalizations of the spurious drifts that appears in ordinary stochastic differential equations with multiplicative noise (see Section~\ref{sec:1D}). Their expression is determined by that of ${\bf C}$; they depend on both time and space discretisations, as explained in Section~\ref{sec:ffpe}. Both obviously vanish when fluctuations are ignored ($T=0$). 

The dynamics~\eqref{eq:dyn_phi} have been used extensively to reproduce the phase separation of active particles~\cite{Speck2013, Tailleur2013, Wittkowski2014, Nardini2017, Nardini2018, Rapp2019}, with~Active Model B+ as a leading example of such theories. In~these works, the~dynamics of the driving chemicals were not considered so that the noise $\boldsymbol\Lambda$ seems to be purely additive. For~this reason, and~because previous studies were not concerned with thermodynamic consistency, the~term $T\boldsymbol\nu$ was missing. Where possible, the~simplest approach to embedding Equation~\eqref{eq:dyn_phi} unchanged within a larger, thermodynamically consistent model is therefore to seek a discretisation scheme (that is, an~interpretation of the original stochastic field theory) in which this spurious drift becomes~zero. 


To date, we did not specify the explicit form of ${\boldsymbol\nu}$ and $\chi$. As~explained above, to~do so requires the discretised version of the dynamics,~(\ref{eq:dyn_phi}) and (\ref{eq:dyn_chem_g}), where we focus on 1D \mbox{for simplicity:}
\begin{equation}\label{eq:dyn_disc}
\begin{aligned}
\dot \phi_i &= \sum_j A_{ij} \Big( \Gamma\sum_k A_{jk}\psi_k - \Delta\mu_j C_j  - T \nu_j - \Lambda_j \Big) \, ,
\\
\dot n_i &= \gamma\Delta\mu_i - C_i \sum_j A_{ij}\psi_j + T \chi_i + \xi_i \, .
\end{aligned}
\end{equation}
Here, $\psi_i = (\partial{\cal F}/\partial\phi_i)/\Delta x$, and~the coupling term $C_i = C(\phi_i, \sum_j A_{ij}\phi_j, ...)$ depends on $\phi$ and its gradients. The~discrete noise terms $\{\Lambda_i,\xi_i\}$ are Gaussian with zero mean and correlations given by
\begin{equation}\label{eq:corr}
\big\langle \big[\Lambda_i,\xi_i\big](t) \big[\Lambda_j,\xi_j\big]^\mathsf{T}(0) \big\rangle = 2 T \,{\mathbb L}_i \frac{\delta_{ij}\delta(t)}{\Delta x} \, ,
\quad
{\mathbb L}_i =
\begin{bmatrix}
\Gamma & C_i
\\
C_i & \gamma
\end{bmatrix}
\, .
\end{equation}
Given that the correlations between $\Lambda_i$ and $\xi_i$ depend on the variable $\phi_i$ through the coupling term $C_i$, one has to specify the temporal discretisation scheme of~Equation \eqref{eq:dyn_disc}. In~what follows, we choose the Stratonovich convention, which allows one to use the standard rules of differential calculus~\cite{Gardiner}. As~found in Section~\ref{sec:Elsen-EP} above, there are compelling reasons to prefer this choice when deriving the expression of the heat rate or~EPR. 

The associated FPE for the probability density $P(\{\phi_i,n_i\},t)$ can then be derived following standard methods~\cite{Gardiner} as
\begin{equation}\label{eq:fpe}
\begin{aligned}
\dot P &= \sum_{i,j} A_{ij} \frac{\partial}{\partial\phi_i} \Big[\Big( - \Gamma \sum_k A_{jk}\psi_k + \Delta\mu_j C_j + T\nu_j \Big) \,P \,\Big] \\
&+ \sum_i \frac{\partial}{\partial n_i} \Big[\Big( - \gamma \Delta\mu_i + C_i \sum_j A_{ij}\psi_j - T \chi_i \Big) \,P\,\Big]
\\
&+ \frac{T}{\Delta x} \sum_{i,a,b,c} \bigg[ \sum_j A_{ij} \frac{\partial}{\partial\phi_j}, \frac{\partial}{\partial n_i}\bigg]_a {\mathbb M}_{i,ab} \bigg[ \sum_k A_{ik} \frac{\partial}{\partial\phi_k}, \frac{\partial}{\partial n_i}\bigg]^\mathsf{T}_c \big( {\mathbb M}_{i,cb} P \big) \, ,
\end{aligned}
\end{equation}
where we have introduced the matrix ${\mathbb M}_i$ defined by ${\mathbb M}_i {\mathbb M}_i^\mathsf{T} = {\mathbb L}_i$. In~the continuum limit of small $\Delta x$, it follows using Equation~\eqref{eq:def-op} that~Equation \eqref{eq:fpe} converges to the standard functional FPE for the probability density $P([\phi(x),n(x)],t)$~\cite{Tailleur2008, Solon2015}. Importantly, by~taking $\{\nu_i,\chi_i\}$ as
\begin{equation}\label{eq:spurious}
\big[\nu_i, \chi_i\big]_a = \frac{1}{\Delta x} \sum_{b,c} {\mathbb M}_{i,ab} \bigg[ \sum_k A_{ik} \frac{\partial}{\partial\phi_k}, \frac{\partial}{\partial n_i}\bigg]_c {\mathbb M}_{i,cb} \, ,
\end{equation}
{the}  
stationary solution of~Equation~\eqref{eq:fpe} is given by the Boltzmann distribution $P_{\rm s}\sim{\rm e}^{- \Delta x \,F/T}$ at equilibrium, namely when $[\psi_i, \Delta\mu_i] = [ \partial F/\partial\phi_i, -\partial F/\partial n_i]$, as~expected~\cite{Chaikin, Lau2007}. As~a result, the~expression of $\{{\mathbb L}_i,\nu_i,\chi_i\}$ in~Equation~\eqref{eq:corr} and~Equation~\eqref{eq:spurious} provides a systematic way to compute the spurious drift terms in terms of $C_i$. When $C_i$ is independent of $n_i$, as~is assumed below,~Equation~\eqref{eq:spurious} vanishes if $C_i$ only depends on $\phi_i$, namely when it is a local function of $\phi$ independent of its gradients. (This is not true of AMB+.) Moreover, the~extension of~Equation~\eqref{eq:spurious} for $d>1$ follows directly by substituting the $d$-dimensional version of the gradient matrix $\bf A$.

When $d=1$, the~chain rule
\begin{equation}
\frac{\partial {\mathbb M}_{i,ab}}{\partial\phi_j} = \frac{\partial {\mathbb M}_{i,ab}}{\partial C_i} \,\frac{\partial C_i}{\partial\phi_j} \, ,
\end{equation}
allows us to simplify~Equation~\eqref{eq:spurious} as
\begin{equation}\label{eq:spurious_g}
\begin{aligned}
\nu_i &= \frac{1}{\Delta x} \bigg( {\mathbb M}_{i,11} \frac{\partial {\mathbb M}_{i,11}}{\partial C_i} + {\mathbb M}_{i,12} \frac{\partial {\mathbb M}_{i,12}}{\partial C_i} \bigg) \sum_j A_{ij} \frac{\partial C_i}{\partial\phi_j} \, ,
\\
\chi_i &= \frac{1}{\Delta x} \bigg( {\mathbb M}_{i,21} \frac{\partial {\mathbb M}_{i,11}}{\partial C_i} + {\mathbb M}_{i,22} \frac{\partial {\mathbb M}_{i,12}}{\partial C_i} \bigg) \sum_j A_{ij} \frac{\partial C_i}{\partial\phi_j} \, .
\end{aligned}
\end{equation}
The matrix ${\mathbb M}_i$ can be written as ${\mathbb M}_i = {\mathbb P}_i^{-1}{\mathbb D}^{1/2}_i{\mathbb P}_i$, where
\begin{equation}
\begin{aligned}
{\mathbb D}_i &=
\begin{bmatrix}
{\tau_{i,-}} & 0
\\
0 & {\tau_{i,+}}
\end{bmatrix}
\, ,
\quad
{\mathbb P}_i =
\begin{bmatrix}
(\tau_{i,-} - \gamma)/C_i & 1
\\
(\tau_{i,+} - \gamma)/C_i & 1
\end{bmatrix}
\, ,
\\
\tau_{i,\pm} &= \frac{1}{2} \bigg[ \gamma+\Gamma \pm \sqrt{4 C_i^2 + (\gamma-\alpha)^2} \bigg] \, .
\end{aligned}
\end{equation}
Substituting the expression of ${\mathbb M}_i$ in~Equation~\eqref{eq:spurious_g}, we find that $\nu_i$ vanishes for any $C_i$ in $d=1$ (it can still potentially be non-zero in higher dimensions), while the expression of $\chi_i$ is
\begin{equation}\label{eq:pi}
\chi_i = \frac{1}{\Delta x} \,\frac{2C_i^2+(\gamma-\Gamma)\Big[\gamma-\sqrt{\gamma\Gamma-C_i^2}\Big]}{4C_i^2+(\gamma-\Gamma)^2} \sum_j A_{ij} \frac{\partial C_i}{\partial\phi_j} \, .
\end{equation}

For the specific coupling term $C_{\rm AMB} \propto \partial_x(\partial_x\phi)^2 = 2 (\partial_x\phi)\partial_{x}^2\phi$ corresponding to Active Model B~\cite{Cates2015} (and in $d=1$ AMB+ also), it is possible to write $C$ using at least two different discretisation schemes, for~example, those used in Equation~\eqref{eq:Cphi-discretised}. The~results in~Equations~(\ref{eq:spur_1} and \ref{eq:spur_2}) are then also appropriate in our case and~illustrate that the choice of spatial discretisation drastically affects the form of the spurious drift terms appearing in the Langevin equations at the field level. Specifically, for~the choice in Equation~(\ref{eq:spur_1}), the spurious drift in Equation~\eqref{eq:pi} vanishes, while for the choice Equation~(\ref{eq:spur_2}), it does not vanish but~instead diverges as $1/\Delta x$. Clearly, therefore, any attempt to numerically code the coupled Langevin \mbox{Equations (\ref{eq:dyn_phi}) and (\ref{eq:dyn_chem_g})} that either ignores the spurious drift terms or~claims to calculate them without reference to the discretisation scheme used risks very large errors in the simulated~dynamics.

\subsubsection{Calculation of the Heat Production~Rate}\label{app:entropy}

We next calculate the heat production rate~\cite{Markovich2021,Seifert2012} 
\begin{equation}\label{eq:entropy}
\dot{\cal Q} = T \left< \underset{t\to\infty}{\lim} \frac{1}{t} \ln \frac{{\cal P}\big[\{{\bf J}, \dot n\}_0^t\big]}{{\cal P}^{\rm R}\big[\{{\bf J}, \dot n\}_0^t\big]} \right> \, ,
\end{equation}
where the average is taken with respect to noise realizations (or ${\cal P}\big[\{{\bf J}, \dot n\}_0^t\big]$). Note that $\dot{\cal Q}/T$ is the {\it full} EPR of our enlarged, thermodynamic model.
The conserved field $\phi$ and its driving $\dot n$ dynamics can be written as
\begin{equation}\label{eq:dyn_gen}
\begin{aligned}
\dot\phi &= -\nabla\cdot{\bf J} \, ,
\\
\begin{bmatrix}
{\bf J} \\ \;\dot n
\end{bmatrix}
&= {\mathbb L}
\begin{bmatrix}
\;-\nabla(\delta{\cal F}/\delta\phi)\; \\ \Delta\mu
\end{bmatrix}
+ T
\begin{bmatrix}
\;{\boldsymbol\nu}\; \\ \chi
\end{bmatrix}
+
\begin{bmatrix}
\;{\boldsymbol\Lambda}\; \\ \xi
\end{bmatrix}
\, ,
\end{aligned}
\end{equation}
where the noise and spurious drift terms obey Equations~\eqref{eq:corr} and \eqref{eq:spurious}, respectively. Generalizing beyond the dynamics in~Equations~(\ref{eq:dyn_phi}) and \eqref{eq:dyn_chem_g}, we now consider an arbitrary Onsager matrix $\mathbb L$, with~the only constraint that it should be positive semi-definite ($\det{\mathbb L}\geq0$).

Following~\cite{Lau2007, Lecomte2017} and similarly to the finite-dimensional case considered in Section~\ref{sec:Elsen-d>1_Action}, the~path probability ${\cal P}\sim{\rm e}^{-{\cal A}}$ associated with~Equation~(\ref{eq:dyn_gen}) is defined by
\begin{equation}\label{eq:action_field}
\begin{aligned}
{\cal A} &= \frac{1}{4T} \int_0^t \int_V
\left(
\begin{bmatrix}
{\bf J} \\ \;\dot n
\end{bmatrix}
+ {\mathbb L}
\begin{bmatrix}
\;\nabla(\delta{\cal F}/\delta\phi)\; \\ -\Delta\mu
\end{bmatrix}
\right)
{\mathbb L}^{-1}
\left(
\begin{bmatrix}
{\bf J} \\ \;\dot n
\end{bmatrix}
+ {\mathbb L}
\begin{bmatrix}
\;\nabla(\delta{\cal F}/\delta\phi)\; \\ -\Delta\mu
\end{bmatrix}
\right)^\mathsf{T}
{\rm d}{\bf r}{\rm d} s \\
&+ {\cal A}_{\rm conv}\, ,
\end{aligned}
\end{equation}
where, as~a consequence of the Stratonovich discretisation, no spurious drift terms appear in the expression~\eqref{eq:action_field}~\cite{Lau2007}. 
Note that, as~before (see Sections~\ref{sec:additive} and \ref{sec:onsager_2d}), ${\cal A}_{\rm conv}$ is even under time-reversal and is not written explicitly in~Equation~\eqref{eq:action_field} since it is not relevant for deriving the heat rate via Equation~(\ref{eq:entropy}). (However, it could potentially be relevant if one or several of the order parameters were odd under time reversal, see,{ e.g.},~\cite{MarLub2021}.) The time-reversed dynamic action ${\cal A}^{\rm R}$ follows from~Equation~\eqref{eq:action_field} by changing the sign of $[{\bf J}, \dot n]$. From~the definition in~Equation~\eqref{eq:entropy}, the~heat rate can be written as
\begin{equation}
\dot{\cal Q} = \lim_{t\to\infty}\frac{T}{t} \left<{\cal A}^{\rm R}-{\cal A}\right> \, ,
\end{equation}
{yielding}  

\begin{equation}\label{eq:entropy_a}
\dot{\cal Q} = \int_V \bigg\langle \bigg\langle \dot n \Delta\mu - {\bf J}\cdot \nabla\frac{\delta\cal F}{\delta\phi} \bigg\rangle \bigg\rangle_t\,{\rm d}{\bf r} \, ,
\end{equation}
where $\lim_{t\to\infty} \frac{1}{t}\int_0^t \cdot \equiv \langle \cdot \rangle_t$ is the steady-state time average. In~steady state, even in spatially inhomogeneous systems, such as phase separation, the~two averages are the same and the temporal one may be omitted.  Note that the product above is interpreted here and in what follows with the Stratonovich~convention.

Integrating by parts the second term in~Equation~\eqref{eq:entropy_a} and using $\dot\phi=-\nabla\cdot{\bf J}$, we obtain $ \int_V \langle {\bf J}\cdot\nabla(\delta{\cal F}/\delta\phi) \rangle {\rm d}{\bf r} = {\rm d}\langle{\cal F}\rangle/{\rm d}t$, which vanishes in steady state, yielding
\begin{equation}\label{eq:entropy_b}
\dot{\cal Q} = \int_V \langle \dot n \Delta\mu\rangle \,{\rm d}{\bf r} \, .
\end{equation}
As a result, the~steady-state heat rate $\dot{\cal Q}$ equals the rate of work injected by the nonequilibrium drive $\Delta\mu$ to sustain the dynamics away from equilibrium. This is equivalent to the first law of thermodynamics, as~expected, when the path probabilities include all thermodynamically relevant fields. For~equilibrium dynamics where $\Delta\mu$ derives from the chemical free energy ${\cal F}_{\rm ch}$, ($\Delta\mu = - \delta{\cal F}_{\rm ch}/\delta n$), 
the heat rate rate vanishes in steady state ($\dot{\cal Q} = -{\rm d}\langle{\cal F}_{\rm ch}\rangle/{\rm d}t = 0$), as~expected. Activity is instead introduced by the fact that $\Delta\mu$ is held away from equilibrium.  (Note that the expression~\eqref{eq:entropy_b} would actually be the same if instead $\dot n$ was held constant and $\Delta\mu$ allowed to fluctuate.)

Substituting the chemical dynamics~\eqref{eq:dyn_chem_g} into~Equation~\eqref{eq:entropy_b}, we deduce
\begin{equation}\label{eq:entropy_c}
\dot{\cal Q} = \gamma V\Delta\mu^2 - \Delta\mu \int_V \bigg\langle {\bf C}\cdot\nabla\frac{\delta\cal F}{\delta\phi} - T \,\chi({\bf C}) \bigg\rangle \,{\rm d}{\bf r} \, .
\end{equation}
Hence, the~heat rate can be separated into (i)~a homogeneous contribution $\gamma V\Delta\mu^2$ corresponding to a background term independent of the fluctuations of the active and chemical fields $\{\phi,n\}$ and~(ii)~a contribution determined only by the fluctuations of the active field $\phi$, with~no contribution from the fluctuations of the chemical coordinate $n$. The~presence of $n$ is nonetheless crucial in determining the form of the heat production rate. Interestingly, the~homogeneous contribution is eliminated when considering  differences in the heat rates at constant $\Delta\mu$, for~example, comparing a state of uniform $\phi$ with a phase-separated one and/or finding the effect on heat rate of changing parameters in the free energy $\cal F$. 

\begin{figure}[h]
	\centering
	\includegraphics[width=0.6\linewidth]{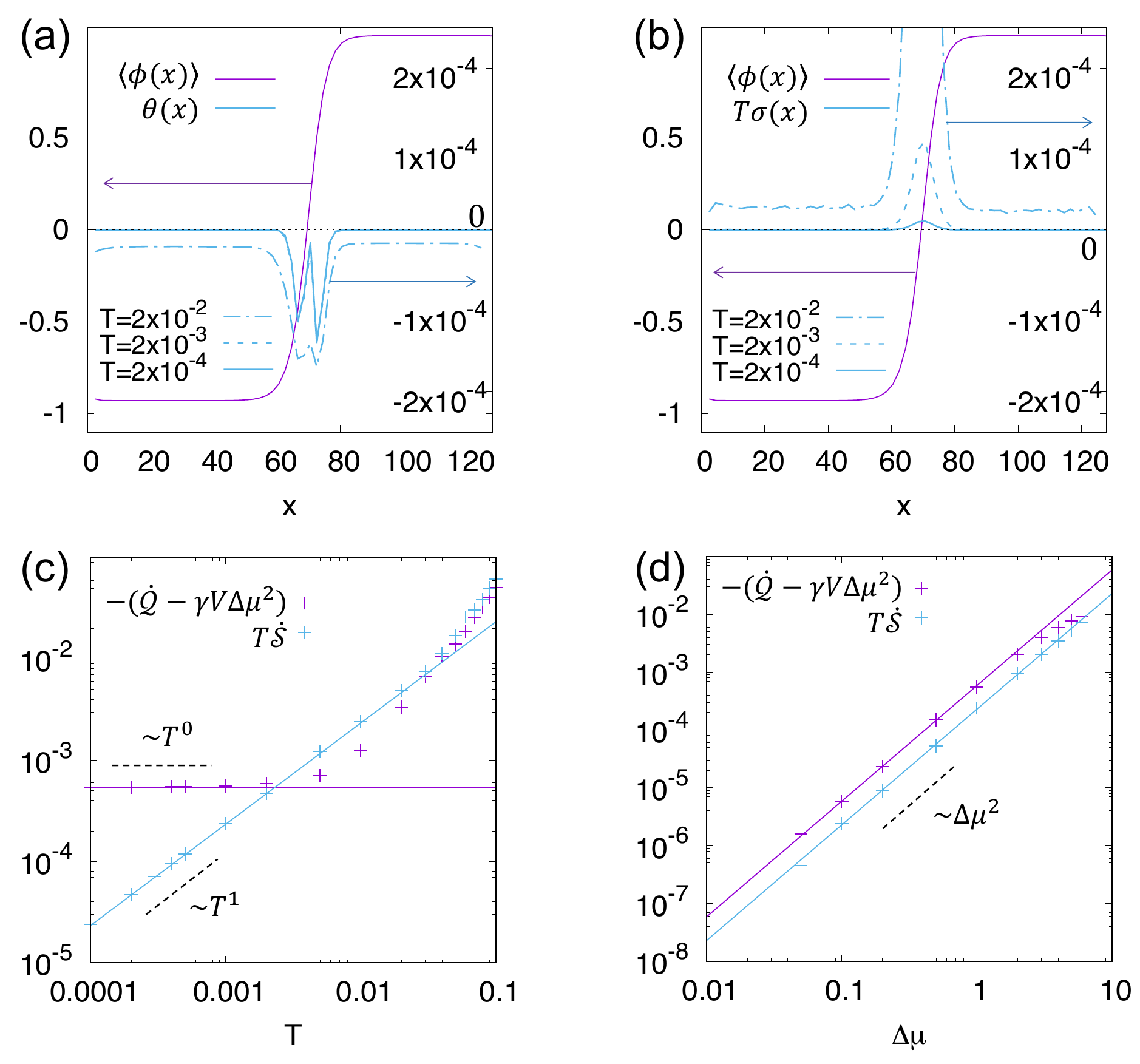}
	\caption{\label{fig:heat_rate}
		Adapted from~\cite{Markovich2021}. Comparison of the heat production rate and IEPR for AMB.
		({\bf a},{\bf b})~The average profile of density $\langle\phi(x)\rangle$ shows a separation between dilute ($\langle\phi(x)\rangle<0$) and dense ($\langle\phi(x)\rangle>0$) phases. The~corresponding profiles of heat rate $\dot q(x)$ and the local IEPR $\sigma(x)$, given, respectively, as: $\dot{\cal Q}-\gamma V\Delta\mu^2 = \int_V \theta  {\rm d}x$ and $\dot{\cal S} = \int_V \sigma {\rm d}x$, are flat in bulk regions and vary rapidly across the interface.
		({\bf c})~The non-trivial contribution to heat rate $\dot{\cal Q}-\gamma V\Delta\mu^2$ reaches a finite value at $T=0$, whereas the IEPR measure $T\dot{\cal S}$ vanishes.
		({\bf d})~$\dot{\cal Q}-\gamma V\Delta\mu^2$ and $T\dot{\cal S}$, respectively, increase and decrease with the driving parameter $\Delta\mu$, and~both scale as $\Delta\mu^2$.
		Parameters used are: $\Gamma=1$, $-a_2=a_4=0.25$, $\kappa=4$, $\bar\phi=0$, $V=128$, 
		$\Delta x = 1$, $\Delta t = 0.01$, (a,b)~$\{\Delta\mu,T\}=\{2,10^{-2}\}$, ({\bf c})~$\Delta\mu=1$, ({\bf d})~$T=10^{-3}$.
	}
\end{figure}

We continue by comparing the heat rate~from Equation~\eqref{eq:entropy_c} with the IEPR as introduced in Section~\ref{sec:additive} and used in previous works~\cite{Nardini2017, Ramaswamy2018, Murrell2019}. Substituting into Equation~\eqref{eq:entropy_c} the expression of $\nabla(\delta{\cal F}/\delta\phi)$ taken from the dynamics~\eqref{eq:dyn_phi} yields
\begin{equation}\label{eq:heat_phi_sd}
\begin{aligned}
\dot{\cal Q} &= T\dot{\cal S} + \frac{\Delta\mu^2}{\lambda} \int_V \big(\lambda\gamma - \big\langle{\bf C}^2\big\rangle\big) \,{\rm d}{\bf r} \\
&+ T \Delta\mu \int_V \left< \chi({\bf C}) - \frac{1}{\lambda} {\bf C} \cdot {\boldsymbol\nu}({\bf C}) -\frac{1}{T\lambda}  {\bf C} \cdot {\boldsymbol \Lambda} \right>   \,{\rm d}{\bf r}  \, ,
\end{aligned}
\end{equation}
where the IEPR  $\dot{\cal S}$ of the $\phi$ dynamics reads~\cite{Nardini2017, Ramaswamy2018, Murrell2019}%
\begin{equation}\label{eq:S_phi}
\dot{\cal S} = \frac{\Delta\mu}{\lambda T} \int_V \big\langle{\bf J}\cdot{\bf C}\big\rangle \,{\rm d}{\bf r} \, .
\end{equation}
(Note that for AMB, this equates by partial integration to Equations~(\ref{eq:mec1a}) and/or (\ref{eq:mec1a-1}) given above.) Clearly, the~second line in Equation~\eqref{eq:heat_phi_sd} depends on the spurious drift terms, but~it also depends directly on the evaluation of the stochastic integral $\int_V\langle{\bf C}\cdot{\boldsymbol\Lambda}\rangle{\rm d}{\bf r}$ and~thereby on the discretisation scheme used to evaluate the heat rate (including spatial discretisation).
We show below that for AMB(+) in $d=1$, a discretisation scheme can be found for which $\{{\boldsymbol\nu},\chi\}=\{{\bf 0},0\}$ and $\int_V\langle{\bf C}\cdot{\boldsymbol\Lambda}\rangle{\rm d}{\bf r}=0$. In~this and other cases for which all these terms vanish, we arrive at a simple relation involving the heat rate $\dot{\cal Q}$ and the \mbox{IEPR, $\dot{\cal S}$:}
\begin{equation}\label{eq:entropy_c_bis}
\begin{aligned}
\dot{\cal Q} &= T\dot{\cal S} + \frac{\Delta\mu^2}{\lambda} \int_V \big(\lambda\gamma - \big\langle{\bf C}^2\big\rangle\big) \,{\rm d}{\bf r}  \,.
\end{aligned}
\end{equation}
From the semi-positivity of the Onsager matrix $\mathbb L$, which ensures $\det{\mathbb L} = \lambda\gamma - {\bf C}^2\geq0$, it then follows that $T\dot{\cal S}$ is a lower bound to $\dot{\cal Q}$. The~bound is saturated when $\bf J$ and $\dot n$ are proportional ($\det{\mathbb L} = 0$): In such a case, the~fluctuations of $\dot n$ are determined by that of $\bf J$, so the irreversibility of the whole dynamics can be found from the trajectories of $\bf J$ alone. As~noted in Section~\ref{sec:additive}, $\dot{\cS}$ can be written as the spatial integral of a local quantity $\sigma(\bfr)$, and~we see that so can be the chemical contribution in Equation~(\ref{eq:entropy_c_bis}). Thus,  $\dot{\cal Q} = \int \dot{q}({\bfr}) \, \D \bfr$ with $\dot{q}$ a local heat production rate density. As~we found with the IEPR, it is interesting to examine where, in~phase-separated system, this density is large or small (see Figure~\ref{fig:heat_rate}).

A specific choice of discretisation for which $\{{\boldsymbol\nu},\chi\}=\{{\bf 0},0\}$ and $\int_V\langle{\bf C}\cdot{\boldsymbol\Lambda}\rangle{\rm d}{\bf r}=0$, such that Equation~\eqref{eq:entropy_c_bis} holds for AMB+ in $d=1$, is that of Equation~(\ref{eq:spur_1}). To~establish this, we evaluate $\sum_i\big\langle C_i\Lambda_i\big\rangle$ transforming it into an It\^{o} product (see Section~\ref{sec:elsen_SI})
\begin{equation}\label{eq:C_lam}
\begin{aligned}
\sum_i\big\langle C_i\Lambda_i\big\rangle	= T \sum_{i,j} &A_{ij} \,\bigg\langle {\mathbb M}_{i,11} \frac{\partial}{\partial\phi_j} \big({\mathbb M}_{i,11} C_i\big)
\, + {\mathbb M}_{i,12} \frac{\partial}{\partial\phi_j} \big({\mathbb M}_{i,12} C_i\big) \bigg\rangle \, ,
\end{aligned}
\end{equation}
where we have used again that $C_i$ is independent of $n_i$. From~Equations~(\ref{eq:pi}), \eqref{eq:heat_phi_sd}, and~(\ref{eq:C_lam}), it follows that the relation between the heat rate and the IEPR depends on $\sum_j A_{ij} (\partial C_i/\partial\phi_j)$, which vanishes for the discretisation of  Equation~(\ref{eq:spur_1}).

In this case, a~direct comparison of the heat-rate with previous results~\cite{Speck2013, Tailleur2013, Wittkowski2014, Nardini2017, Nardini2018, Rapp2019}, and~specifically with the results of Section~\ref{sec:additive}, which did not have spurious drift terms is valuable. In~Figure~\ref{fig:heat_rate}, we provide such a comparison. 
For a phase-separated profile, as~shown in Figure~\ref{fig:heat_rate}(a-b), the~leading order of $\dot{\cal Q}-\gamma V\Delta\mu^2$ scales like $T^0$, and~it reaches a finite value at $T=0$. Hence, the~heat rate $\dot{\cal Q}$ is not only determined by the background term $\gamma V\Delta\mu^2$ at zero temperature; it now also depends on the mean-field density profile. In~contrast, $T\dot{\cal S}$ scales like $T$ and thus vanishes at $T=0$, see Figure~\ref{fig:heat_rate}c, as~already reported in~\cite{Nardini2017} and in Figure~\ref{fig:sigmarealspace}. Notably, while the IEPR is maximal on the interface between phases, showing maximal irreversibility of the fluctuating $\phi$ dynamics, the~heat rate density is suppressed there. This suggests that the chemical reactions are, in~the interfacial zone, producing less heat because they are instead doing local work against ${\cF}$ to sustain the nonequilibrium coexistence. Thus, both $\dot{\cS}$ and $\dot{\cal Q}$ can differently reveal useful insights into the dynamics of the system.
%

These results also fully confirm that the IEPR, which considers the irreversibility of the $\phi$ dynamics alone, does not capture the full energetic cost of creating phase separation away from equilibrium, as~the heat-rate $\dot{\cal Q}$ does. In~fact, if~$T\dot{\cal S}$ was indeed a measure of the full energetic cost, a~nonequilibrium active phase separation could be sustained at zero energy cost as $T\to 0$, contradicting the basic thermodynamic notion that activity is powered by constant input energy that is ultimately dissipated as~heat.

\unskip

\section{Concluding~Remarks \label{sec:conclusion}}
In this paper, we have addressed several conceptual issues arising from the stochastic PDEs (SPDEs) used by physicists to describe the fluctuating hydrodynamics of complex fluids. These conceptual issues arise because the continuum limit, while implicit in the notation used to write down these SPDEs, is generally either nonexistent or at least problematic~\cite{Hairer}. The~usual physicist's defence is to protest that there is always a short-scale cutoff (set by molecular physics), so the SPDEs are really only a short-hand for a discretised version of the same equations. Rarely are such versions closely examined, and~often, they are not even specified unless numerical work is actually undertaken (sometimes, not even then).  We hope to have convinced the reader that a more careful study of the meaning of these equations based on careful and consistent discretisation strategies \mbox{is warranted. }

In particular, attention must be paid to achieving detailed balance at the discrete level in the case of equilibrium systems. This is not a new remark (see, e.g.,~\cite{RonojoyEPL}) but is brought into sharper focus by the desire to numerically evaluate the entropy production rate (EPR). This desire, driven by recent work on active rather than equilibrium complex fluids, requires careful study of the discretisation scheme used to establish the path weights (or dynamical action), from which, via the laws of stochastic thermodynamics, the~EPR can be calculated. 
Although the scheme to embed SPDEs within thermodynamically consistent description is not unique a priori, our framework provides a minimal approach to do so without LIT. 
Interestingly, LIT is also the starting point for a large class of active field theories, known as active gels, which have been extremely successful in capturing the dynamics of complex biological systems, such as acto-myosin networks and living tissues~\cite{Prost2017,Markovich2019b,Furthauer2012,Kruse2004}.

For SPDEs with purely additive noise (such as Model B and its active variants), these problems are first encountered in computing the {\em informatic} EPR (IEPR) for fluctuating active fields, which quantifies the  irreversibility of the coarse-grained order parameter dynamics without concern for the underlying heat flows. However, the same problems are accentuated further when one addresses these heat flows by minimally coupling the active order parameter fields to an underlying chemical process governed by linear irreversible thermodynamics. In~this case, the~active terms in the stochastic hydrodynamic equations for the order parameters become off-diagonal Onsager couplings in the enlarged model. The~result is that the off-diagonal noise is multiplicative, even when the original noise in the order parameter sector was not. This necessitates the treatment of spurious drift terms directly in the Langevin dynamics; like similar terms in the dynamical action, these are dependent on both temporal and spatial discretisation schemes. Moreover, unless~they can be eliminated altogether by careful design of such schemes, these terms {\em diverge} in the spatial continuum limit, $\Delta x \to 0$.  In~this setting, and~presumably also in other models of fluctuating complex fluids that involve multiplicative noise (for example, Model B with a composition-dependent mobility), relatively minor oversights in numerical implementation could therefore lead to errors in the generation of Langevin trajectories that are not merely ${\cal O}(1)$, but~unbounded, as~the continuum limit is~approached.

\vspace{6pt}

\acknowledgments{We thank Yongjoo Baek, \O yvind Borthne, Patrick Pietzonka, Frederic van Wijland, and~Julien Tailleur for valuable~discussions. This work received funding from the European Research
	Council (ERC) under the EU's Horizon 2020 Programme,
	Grant agreement Nos. 740269 and 885146. CN acknowledges the support of an Aide Investissements d'Avenir du LabEx PALM (ANR-10-LABX-0039-PALM). \'EF
	acknowledges support from an ATTRACT Grant of the
	Luxembourg National Research Fund. 
	TM acknowledges support from the National Science Foundation Center for Theoretical Biological Physics (Grant PHY-2019745).
	MEC is funded
	by the Royal~Society.}

\appendix
\section{Spurious Drift in Linear Irreversible~Thermodynamics}\label{app:sd_lit}

Within the framework of LIT, the~thermodynamic fluxes are written as a linear combination of the thermodynamic forces~\cite{Joanny2009,Marchetti2013,Markovich2019b,Mazur}. The~matrix connecting these is called the Onsager matrix, which must obey Onsager reciprocity relations and Curie's symmetry principle. Notably, there is no other restriction on the Onsager matrix. It can therefore be nonlinear in the fields. 
A generic equation for the flux of some order parameters $\{\psi_a\}$\mbox{ is then:}
\begin{eqnarray}
\label{sd4aa}
\frac{\D\psi_a(\vecr;t)}{\D t} = -\int\D\vecr' {\cal M}_{ab}(\vecr,\vecr',[\phi];t) \frac{\delta {\cal F}\left[ \left\{ \psi_a \right\}\right]}{\delta\psi_b(\vecr')} + \nu_a(\vecr;t) + g_{ab}(\vecr;t) \xi_b(t) \, ,
\end{eqnarray}
where ${\bm \nu}$ is the spurious drift and ${\bm \xi}$ is a Gaussian white noise with variance $\langle \xi_a(\vecr;t) \xi_b(\vecr';t') \rangle = \delta_{ab} \delta(\vecr-\vecr')\delta(t-t')$. Here and below, we separate spatial and time variables with a semicolon. Note that the variable-diffusivity ideal gas equation considered in Section~\ref{sec:ffpe} is a simple example of this type. Because~the noise is multiplicative, the~Langevin equation for the flux is not well defined without specifying how the noise is evaluated (equivalently, specifying a discretisation scheme). As~shown in the main text, for~our purposes, the Stratonovich convention is the most useful and will be used throughout this Appendix. 

Following the derivation in Ref.~\cite{Lau2007}, and~extending it to fields (see also~\cite{Chaikin,Basu2008}), we write the functional FPE
\begin{eqnarray}
\label{csd1b}
\nonumber& &\frac{\partial P}{\partial t} = - \int\D\vecr \frac{\delta J_a\left[ \left\{ \psi_a \right\}\right]}{\delta\psi_a(\vecr)} \, ,
\end{eqnarray}
with
\begin{align}
\label{csd1c}
\nonumber J_a\left[ \left\{ \psi_a \right\}\right] &= \int\D\vecr' \Bigg( \Bigg[ {\cal M}_{ab}(\vecr,\vecr';t){f}_b(\vecr';t) 
+ T \int\D\vecr'' {\mathbb M}_{cb}(\vecr',\vecr'';t)\frac{\delta {\mathbb M}_{ab}(\vecr,\vecr'';t)}{\delta\psi_c(\vecr')} \\
&- \half \frac{\delta \Gamma_{ab}(\vecr,\vecr';t)}{\delta\psi_b(\vecr')} \Bigg]  P 
-\half \Gamma_{ab}(\vecr,\vecr';t) \frac{\delta  P}{\delta\psi_b(\vecr')} \Bigg) \, ,
\end{align}
where $\Gamma_{ab}(\vecr,\vecr';t) \equiv 2T \int\D\vecr'' {\mathbb M}_{ac}(\vecr,\vecr'';t) {\mathbb M}_{bc}(\vecr',\vecr'';t)$, and we define ${\bf f}(\vecr)$ similarly to $f(x)$ of Equation 
~\eqref{eq:1D-Langevin}: 
\begin{eqnarray}
\label{csd1d}
\int\D\vecr'{\cal M}_{ab}(\vecr,\vecr';t){f}_b(\vecr';t) \equiv - \int\D\vecr' {\cal M}_{ab}(\vecr,\vecr';t) \frac{\delta {\cal F}}{\delta\psi_b(\vecr')} + \nu_a(\vecr;t) \, .
\end{eqnarray}

Requiring that the stationary solution of the functional FPE, Equation~(\ref{csd1b}), is $P_{\rm s} \sim \exp\left[ -{\cal F}/T \right]$, the~flux $J_a$ must have the form~\cite{Basu2008,Chaikin}:
\begin{eqnarray}
\label{csd3a}
J^*_a = \int\D\vecr' \Bigg( \Big[ - {\cal M}_{ab}(\vecr,\vecr';t) \frac{\delta F\left[ \left\{ \psi_a \right\}\right]}{\delta\psi_b(\vecr')} + T\frac{\delta {\cal M}^a_{ab}(\vecr,\vecr';t)}{\delta\psi_b(\vecr')} \Big]  P - T {\cal M}_{ab}^s(\vecr,\vecr';t) \frac{\delta  P}{\delta\psi_b(\vecr')} \Bigg) \, .
\end{eqnarray}
This yields the following relations:
\begin{eqnarray}
\label{csd3b}
& &2T {\cal M}_{ab}^s (\vecr,\vecr';t) = \Gamma_{ab}(\vecr,\vecr';t) \, , \\
\label{csd3c}
\nonumber& &{\cal M}_{ab}(\vecr,\vecr';t){f}_b(\vecr') = -{\cal M}_{ab}(\vecr,\vecr';t) \frac{\delta {\cal F}}{\delta\psi_b(\vecr')} 
+ T\frac{\delta {\cal M}^a_{ab}(\vecr,\vecr';t)}{\delta\psi_b(\vecr')} 
+\half  \frac{\delta \Gamma_{ab}(\vecr,\vecr';t)}{\delta\psi_b(\vecr')} \\
& &\qquad\qquad\qquad\qquad- \, T \int\D\vecr'' {\mathbb M}_{cb}(\vecr',\vecr'';t)\frac{\delta {\mathbb M}_{ab}(\vecr,\vecr'';t)}{\delta\psi_c(\vecr')} \, .
\end{eqnarray}
Using Equation~\eqref{csd1d} and substituting Equation~(\ref{csd3b}) into Equation~(\ref{csd3c}), we finally obtain the spurious drift in the Stratonovich convention, 
\begin{eqnarray}
\label{sd4a}
\nu_a(\vecr;t) = T \int\D\vecr'\left[  \frac{\delta {\cal M}^a_{ab}(\vecr,\vecr';t)}{\delta\phi_b(\vecr')} 
+  \int\D\vecr'' {\mathbb M}_{ac}(\vecr,\vecr'';t)\frac{\delta {\mathbb M}_{bc}(\vecr',\vecr'';t)}{\delta\phi_b(\vecr')} \right]    \, .
\end{eqnarray}

Note that ${\mathbb M}$ can be defined as symmetric, in~which case, it is the square root of $\boldsymbol{ M}^s$~\cite{Lau2007}. Note also that the choice of different time-discretisation schemes only affects the dissipative part of the generalized mobility matrix (its symmetric part $\boldsymbol{ M}^s$), while its reactive part $\boldsymbol{ M}^a$ (the antisymmetric part)~\cite{Basu2008,Markovich2019b} contributes a term that is unaffected by~time-discretisation.

Importantly, and~as explained in detail in Sections~\ref{sec:t-discr} and \ref{sec:ffpe}, there are problems with the continuous description of the spurious drift, specifically in cases where $\boldsymbol{ M}$ involves spatial gradients. Therefore, to~make sense of the expressions in this Appendix, they must be discretised. 
Following principles already laid down in the main text, we discretise space and write $\psi_a(\vecr) \to \psi_{a;(i,j,k)}$ with $\{a,b,c\}$ letters denoting the various fields and $\{i,j,k\}$ referring to the spatial discretisation $\vecr \to (i\Delta x, j\Delta y, k\Delta z)$. Then, the~spatially discretised spurious drift in the Stratonovich convention is written as
\begin{eqnarray}
\label{sd4}
\nu_{a;(i,j,k)} = T \sum_{i'j'k'} \Bigg[ \frac{\partial {\cal M}^a_{ab;(i,j,k),(i',j',k')}}{\partial \psi_{b;(i',j',k')}} 
+  \sum_{i''j'k''}   {\mathbb M}_{ac;(i,j,k),(i'',j'',k'')} \frac{\partial {\mathbb M}_{bc;(i',j',k'),(i'',j'',k'')} } {\partial\psi_{c;(i',j',k')}}  \Bigg] \, ,
\end{eqnarray}
where we suppress the time dependence. 

	
	


\end{document}